\documentclass[aps,twocolumn,showpacs,superscriptaddress]{revtex4}

\usepackage{graphicx}  
\usepackage{amsmath}   
\usepackage{amssymb}   
\usepackage{color}
\usepackage[T1]{fontenc} 
\usepackage[dvipsnames]{xcolor}
\usepackage{hyperref}  
\hypersetup{
    colorlinks=true,
    linkcolor=blue,
    filecolor=blue,
    urlcolor=blue,
    citecolor=blue,
    pdfborder={0 0 0}
}

\newcommand{\apjs}{ApJS}

\newcommand{\apjl}{ApJL}
\newcommand{\aap}{A$\&$A}
\newcommand{\mnras}{MNRAS}

\begin{document}
\title{Inferring the  pair-instability mass gap from gravitational wave data}

\author{Fabio Antonini}
\affiliation{Gravity Exploration Institute, School of Physics and Astronomy, Cardiff University, Cardiff, CF24 3AA, UK}
\author{Thomas Callister}
\affiliation{Kavli Institute for Cosmological Physics, The University of Chicago, Chicago, IL 60637, USA}
\author{Fani Dosopoulou}
\affiliation{Gravity Exploration Institute, School of Physics and Astronomy, Cardiff University, Cardiff, CF24 3AA, UK}
\author{Isobel M. Romero-Shaw} 
\affiliation{DAMTP, Centre for Mathematical Sciences, University of Cambridge, Wilberforce Road, Cambridge, CB3 0WA, UK}
\affiliation{Kavli Institute for Cosmology, Madingley Road, Cambridge, CB3 0HA, United Kingdom}
\affiliation{H. H. Wills Physics Laboratory, Tyndall Avenue, Bristol BS8 1TL, UK}
 \author{Debatri Chattopadhyay} 
\affiliation{Center for Interdisciplinary Exploration and Research in Astrophysics (CIERA) and Department of Physics \& Astronomy, Northwestern University, 1800 Sherman Ave, Evanston, IL 60201, USA}

\date{\today}

\begin{abstract}
We use hierarchical Bayesian inference with non-parametric Gaussian process models to investigate the effective inspiral spin parameter, $\chi_{\rm eff}$, as a function of primary black hole mass in the  third gravitational-wave transient catalog
 (GWTC-3).  
Our analysis reveals a transition in the population
at a primary mass of $46^{+7}_{-5}\,M_\odot$. Beyond this mass, the $\chi_{\rm eff}$ distribution broadens, becomes consistent with being symmetric around zero, and has a median of $-0.03^{+0.36}_{-0.59}$ (90\% credibility).
These results are consistent with the presence of a pair-instability mass gap that is  repopulated by black holes that are the remnant of a previous merger, formed in dense star clusters.
However, asymmetric distributions skewed toward positive $\chi_{\rm eff}$ are not excluded by current data.
Below the inferred transition mass, we constrain the fraction of second-generation black holes to be  $\lesssim 10\%$.
These results provide model-independent support for a  high-mass and high-spin  population of  black holes in the data, consistent with earlier work using parametric models.
Imminent gravitational-wave data releases will be essential to sharpen constraints on spin symmetry and clarify the origin of the black holes.
\end{abstract}


\maketitle


\section{Introduction}
Observations of gravitational waves (GWs) from binary black hole (BH) mergers have revolutionized our ability to probe the lives and deaths of massive stars \cite{2019PhRvX...9c1040A, 2019ApJ...882L..24A, Abbott:2020gyp, 2021arXiv211103606T, 2021PhRvX..11b1053A}. These mergers encode critical information about the formation, evolution, and final fates of massive binary systems across cosmic time. However, interpreting the growing sample of detections remains challenging, largely due to significant uncertainties in the physics of binary stellar evolution and BH formation. These include poorly understood processes such as mass transfer, common-envelope evolution, natal kicks, and spin alignment, as well as the sensitivity of population synthesis models to uncertain initial conditions and parameter choices \cite[e.g.,][]{Spera2015a, belczynski2016, Stevenson2015, Marchant2016}.

A key open question is the existence and extent of a predicted mass gap in the BH birth mass distribution, caused by (pulsational) pair-instability supernovae ((P)PISNe). 
Although the location of the mass gap is subject to several uncertainties \cite[e.g.,][]{2021MNRAS.502L..40F,2020ApJ...890..113B},
stellar evolution theory predicts that stars with helium core masses in the range $\approx 40$–$65\,M_\odot$ undergo partial mass ejection (pulsational PISN), while those above \(\sim 65\,M_\odot\) are entirely disrupted (full PISN), preventing the formation of BHs in the approximate range $\sim40$–$130\,M_\odot$ \cite{Woosley2016, 2017MNRAS.470.4739S, 2020ApJ...902L..36F,2023MNRAS.526.4130H,2019ApJ...887...53F,2019ApJ...887...72L}. Yet, current GW observations from the LIGO-Virgo-KAGRA Collaboration (LVK) show no sharp cutoff or dearth of mergers in this mass range \cite{2019ApJ...882L..24A,2021ApJ...913L...7A,2021ApJ...913L..23E,LVKCollab2023,ray_nonparametric_2023,2024PhRvX..14b1005C,2023arXiv230302973L}, raising questions about either the location of the gap or the mechanisms responsible for populating it
\cite[e.g., ][]{2021MNRAS.502L..40F,2020ApJ...902L..36F,2024MNRAS.529.2980W}.

One compelling explanation is the formation of high mass BHs from previous mergers~\footnote{In the literature, such events are often termed "hierarchical mergers," implying growth through multiple successive mergers. However, we avoid this terminology, as theoretical models predict that the vast majority of BHs in the mass gap originate from a single prior merger—i.e., they are predominantly second-generation BHs.}. Stellar-origin BHs undergo successive mergers in dense stellar environments, such as globular clusters, nuclear star clusters, or 
active galactic nucleus (AGN) disks \cite{2006ApJ...637..937O,2016ApJ...831..187A,Rodriguez2015a, 2017PhRvD..95l4046G, 2019PhRvL.123r1101Y, 2019MNRAS.486.5008A, 2019MNRAS.486.4781F, 2021MNRAS.508.3045D, 2021MNRAS.505..339M, 2021MNRAS.507.3362T}. In these scenarios, first-generation BHs merge to form more massive, second-generation BHs, which can then participate in further mergers. This process naturally populates the pair-instability mass gap and leads to distinctive spin signatures. The BH components involved in such  mergers are expected to have higher spins and isotropic spin orientations \cite{2017ApJ...840L..24F, 2017PhRvD..95l4046G, 2020PhRvD.102d3002B}. Detecting such a subpopulation would not only confirm the presence of the (P)PISN mass gap but also offer a direct probe of dynamical processes in star clusters and galactic nuclei.

The most informative spin parameter for detecting such effects is the effective inspiral spin, \(\chi_{\rm eff}\), defined as the mass-weighted projection of the component BH spins onto the orbital angular momentum \cite{ajith2011, Vitale}. \(\chi_{\rm eff}\) is relatively well constrained in GW signals and serves as a key discriminator between isolated and dynamical formation channels. Isolated binaries formed through common-envelope evolution are expected to have low and moderately aligned spins, leading to positive-skewed \(\chi_{\rm eff}\) distributions \cite{farr_distinguishing_2017,2023PhRvD.108j3009G}, while dynamically assembled binaries are expected to produce broader, symmetric \(\chi_{\rm eff}\) distributions centred around zero~\cite{2017CQGra..34cLT01V,2017PhRvD..95l4046G,Rodriguez2016c}. Additional parameters—such as the precessing spin parameter \(\chi_p\) \cite{schmidt2015}, the mass ratio \(q = m_2/m_1\), the orbital eccentricity, and the redshift \(z\)—have also been leveraged to identify subpopulations and trends indicative of dynamical formation \cite{2021ApJ...910..152Z, 2021NatAs...5..749G, 2021ApJ...922L...5C, 2022ApJ...937L..13C, 2022arXiv220614695R,2022ApJ...928...75H,2025PhRvD.111b3037H}. In recent years, several studies have investigated correlations between mass and spin as a signature of dynamical mergers \cite{2020ApJ...900..177K, 2021MNRAS.502.2049L, 2022PhRvD.105l3024F, 2023arXiv230302973L,2024arXiv240601679P,2025PhRvD.111f3043H,2024arXiv240616844H}.

A key prediction of the merger scenario in dense clusters is the emergence of a distinct population of second-generation BHs with isotropic spin orientations and well-defined spin magnitude boundaries. In particular, if the pair-instability mass gap is populated by BHs formed from a previous merger, the resulting effective spin distribution is expected to be broad, symmetric about zero, and truncated at characteristic limits set by the spin of the merger remnant—typically $a\simeq 0.7$, where $a$ is the dimensionless spin parameter of the BH, $a=cj/Gm^2$ with $j$ the spin angular momentum. This feature arises from robust physical considerations and is largely independent of the specific properties of the host cluster. 
In Ref.~\cite{2025PhRvL.134a1401A}, 
 we showed that the $ \chi_{\rm eff}$ 
distribution of BH binaries in which the primary was formed from a previous merger is
a uniform distribution with boundaries at $|\chi_{\rm eff}| \simeq 0.47$. This corresponds to the cumulative distribution function ($\mathrm{CDF}$):
\begin{equation}\label{CDFeff}
\mathrm{CDF}(\leq \chi_{\rm eff}) \simeq 0.5 +{\chi_{\rm eff}}~.
\end{equation} 
We note that Refs.~\cite{2020PhRvD.102d3002B, 2021PhRvD.104h4002B} obtained a similar result, but with a significantly smaller
boundary of the uniform distribution $|\chi_{\rm eff}| \simeq 0.3$. In Ref.~\cite{2025PhRvL.134a1401A} we showed that a larger $|\chi_{\rm eff}| \simeq 0.47$ is 
required to match the results of cluster simulation models and theoretical expectations.

Other formation channels, such as gas-assisted mergers in AGN discs,  or field evolution of triples or higher multiplicity systems, can, in principle, also produce high-mass systems in the mass gap \cite[e.g.,][]{2020ApJ...895L..15F,2021ApJ...908..194T}. However, these channels do not yield similarly robust or universal predictions for the \(\chi_{\rm eff}\) distribution. For example, the degree of spin alignment in AGN disks is sensitive to uncertain details of disk structure, migration, and accretion efficiency, while spin outcomes in isolated multiple star systems depend heavily on natal spin distributions, tides, and stellar evolution prescriptions. As a result, while our analysis —focused on identifying clear, data-driven signatures in the \(\chi_{\rm eff}\) distribution— can  be used  to test the cluster formation hypothesis, it provides limited discriminatory power for other formation channels with poorly-constrained spin predictions.

In Ref.~\cite{2025PhRvL.134a1401A}, we developed a parametric mixture model for the effective spin distribution. In that study, the \(\chi_{\rm eff}\) distribution was modeled as a mass-dependent combination of two populations: a low-spin Gaussian component, and a broad, near-uniform component representing the spin distribution expected from dynamically assembled, second-generation BHs. This analysis found strong evidence for a high-mass component with a broad, near-isotropic spin distribution becoming dominant above \(\simeq 45\,M_\odot\), consistent with the onset of the (P)PISN mass gap. However, the extent to which these conclusions depend on the specific parametric form of the model remained an open question.

In this work, we 
consider hierarchical inference of the effective spin distribution of BH binaries
 using data from the third gravitational-wave transient catalogue~\cite[GWTC-3,][]{2021arXiv211103634T}, employing flexible, non-parametric models that make minimal assumptions about the underlying population. Our primary aim is to determine the mass scale at which a subpopulation with spin properties consistent with 2nd generation BHs begins to emerge and to evaluate the statistical significance and astrophysical interpretation of this transition. By modeling both the spin and mass distributions non-parametrically, we allow the data to guide the inference without imposing strong theoretical priors, enabling a more robust and data-driven exploration of spin–mass correlations. We find consistent a transition mass of $\approx45\,M_\odot$, consistent with our earlier study.
 
The rest of this paper is organized as follows.
In Sec.~\ref{sec:methods} we describe our methods and model used.
In Sec.~\ref{sec:results} we present our results, describing observed trends in BH effective spins and mass ratios as a function of primary mass.
In Sec.~\ref{sec:astro}, we comment on possible astrophysical interpretations of these results, and conclude in Sec.~\ref{sec:conclusions}. In Appendices~\ref{mass} and \ref{qvsm} we report on the inferred primary mass distribution and present an additional model where we model the mass-ratio distribution as a function of primary mass.
More details about the data and the methodology,  as well all priors on the hyperparameters of our models are reported in Appendices~\ref{app:inference-methods} and \ref{priors}.

\section{Data and baseline model}
\label{sec:methods}
In the analysis that follows, we consider the subset of binary BH (BBH) mergers in GWTC-3 with false alarm rates below \(1\,{\rm yr}^{-1}\), consistent with Ref.~\cite{LVKCollab2023}. We exclude all events  which include at least one component with mass \(< 3\,M_\odot\) and is therefore likely to involve a neutron star~\cite{LVKCollab2023}. The resulting dataset contains 69 BBH mergers.
Selection effects are accounted for using the set of successfully recovered BBH injections made publicly available by the LIGO–Virgo–KAGRA collaboration, covering their first three observing runs~\cite{LVKCollab2023,injections}; see Appendix~\ref{app:inference-methods} for further details.

We assume the merger rate density can be factorized as:
\begin{equation}
\begin{aligned}
R(m_1, m_2, \chi_{\rm eff}; z) &= 
    R_{\rm ref}\,\frac{f(m_1)}{f(20\,M_\odot)} \left(\frac{1+z}{1.2}\right)^\kappa \\
    &\qquad \times p(m_2 | m_1)\, p(\chi_{\rm eff} | m_1),
\end{aligned}
\end{equation}
where \(R_{\rm ref}\) denotes the merger rate per unit mass evaluated at redshift \(z = 0.2\) and primary mass \(m_1 = 20\,M_\odot\).
Our primary object of study is $p(\chi_\mathrm{eff}|m_1)$, the probability distribution of effective spins as a function of BH primary mass.
In Sec.~\ref{sec:results} below, we will adopt several different models for the probability distribution $p(\chi_\mathrm{eff}|m_1)$, ranging from parametric to flexible non-parametric approaches.

We will simultaneously infer the distributions of BBH primary masses, mass ratios $q$, and redshifts $z$.
Unless otherwise specified, we model the primary mass distribution non-parametrically via a Gaussian process (GP).
In particular, we take
\begin{equation}\label{massGP}
f(m_1) = \mathrm{exp}[\Phi(\ln m_1)]
\end{equation}
where $\Phi(\ln m_1)$ is drawn from a zero-mean GP prior,
\begin{equation}
\Phi(x) \sim \mathcal{GP}\left( 0,\, k(x, x';\, a_m, \ell_m) \right)
\end{equation}
with a squared-exponential kernel
\begin{equation}
\label{eq:kernel}
k(x, x') = a_m^2 \exp\left( -\frac{(x - x')^2}{2\ell_m^2} \right).
\end{equation}
Here, \(a_m\) is the amplitude of the GP (controlling vertical variation), and \(\ell_m\) is the length scale (controlling smoothness). We treat both \(a_m\) and \(\ell_m\) as free hyperparameters, assigning priors that allow for flexibility while ensuring the inference remains robust and the effective sample size remains sufficient (see the Appendix).

In practice, we evaluate the GP on a uniform grid in \(\log m_1\) over the range \(2\,M_\odot\) to \(100\,M_\odot\), using grid points \(\{x_i = \log m_1^{(i)}\}\), for \(i = 1, \ldots, N_{\rm bin}\), with \(N_{\rm bin} = 100\). The latent function values \(\mathbf{y} = [\Phi(x_1), \ldots, \Phi(x_{N_{\rm bin}})]^\top\) are then drawn from the multivariate normal distribution:
\[
\mathbf{y} \sim \mathcal{N}(\mathbf{0},\, \mathbf{K}),
\]
where the covariance matrix is defined by \(\mathbf{K}_{ij} = k(x_i, x_j; a_m, \ell_m)\). Samples are generated via Cholesky decomposition of \(\mathbf{K}\) to ensure numerical stability.
The values of $\mathbf{y}$ are then interpolated to evaluate the GP at the locations of events' posterior samples and recovered injections.

We model the conditional distribution of the secondary mass \(m_2\) following Ref.~\cite{2022ApJ...937L..13C}:
\begin{equation}
\label{eq:pm2}
p(m_2 | m_1) \propto m_2^{\beta_q}, \qquad 2\,M_\odot \leq m_2 \leq m_1.
\end{equation}
Meanwhile, we assume that the volumetric merger rate evolves as a power law in $(1+z)$~\cite{Fishbach_2018,Callister_2020}, such that probability distribution of merger redshifts is
\begin{equation}
\label{eq:pz}
p(z) \propto \frac{1}{1+z} \frac{dV_c}{dz} (1+z)^\kappa.
\end{equation}

\begin{figure} 
\includegraphics[width=1\columnwidth]{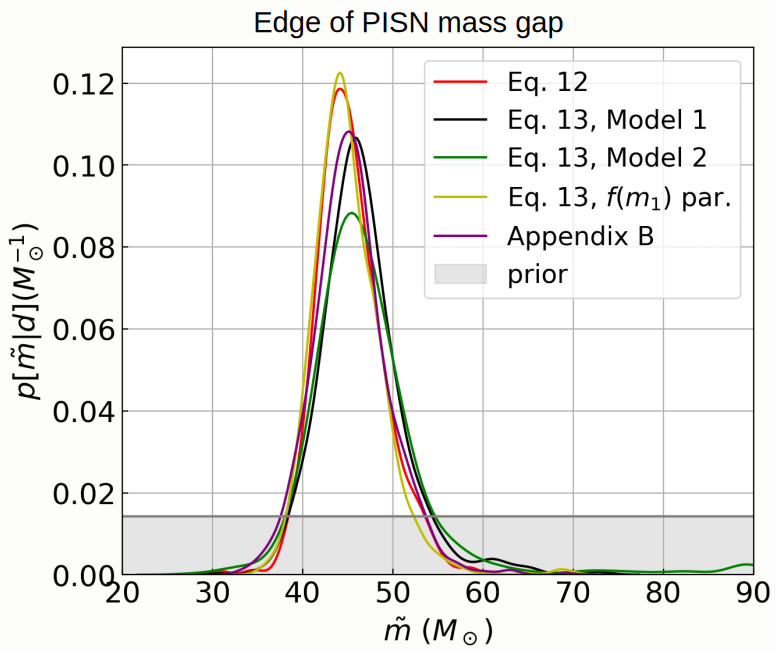} 
    \caption{ Posteriors of the transitional mass $\tilde{m}$ obtained under the different models considered in this work.
    All models  yield a consistent estimate of $\tilde{m} \simeq 45\,M_\odot$
    The black and purple lines correspond to models in which both the primary mass distribution and the $\chi_{\rm eff}$ distribution are described non-parametrically; these differ only in their prior choices, referred to as Model 1 and Model 2 in the main text. 
The yellow line corresponds to a model in which the mass distribution is represented parametrically using the standard power-law plus peak prescription commonly used in previous studies. The red line corresponds to a model in which the $\chi_{\rm eff}$ distribution is represented by the parametric form given in equation~(\ref{UL_ind}), i.e., a uniform distribution with independent bounds. 
Finally, the purple line denotes a model in which the mass ratio power-law slope parameter $\beta_q$ is modeled as a GP over the primary BH mass (see Appendix~\ref{qvsm}). 
}
    \label{fig:tmass}
\end{figure}

\section{Results}
\label{sec:results}
We begin by highlighting a key result of this study: the primary mass scale \(\tilde{m}\) at which BH spins transition, previously inferred using strongly parametric models as \(\tilde{m} = 44^{+6}_{-4}\,M_\odot\)~\cite{2025PhRvL.134a1401A}, is recovered here using significantly more flexible, non-parametric models. In Ref.~\cite{2025PhRvL.134a1401A}, the distribution of \(\chi_{\rm eff}\) was modeled as a mixture of a Gaussian component, representing the bulk population at \(m_1 \lesssim \tilde{m}\), and a uniform component, associated with dynamically assembled (1st+2nd generation BH)  mergers at \(m_1 \gtrsim \tilde{m}\).
The parameter \(\tilde{m}\) denotes the transition mass where the dominant formation channel shifts from isolated to dynamical.

Figure~\ref{fig:tmass} displays the inferred posterior distribution of \(\tilde{m}\) for all models considered in this work. Across the different model classes, to be discussed further below, we consistently recover a transition in spin properties at a characteristic primary mass of \(\tilde{m} = 46^{+7}_{-5}\,M_\odot\) (90\% credible interval). Above this mass, we find strong evidence for the emergence of a distinct  subpopulation of BHs with systematically higher spins.

In the sections that follow, we describe the individual models in detail and examine their respective predictions for the mass–spin correlation.

\begin{figure}    
\includegraphics[width=1\columnwidth,angle=0]{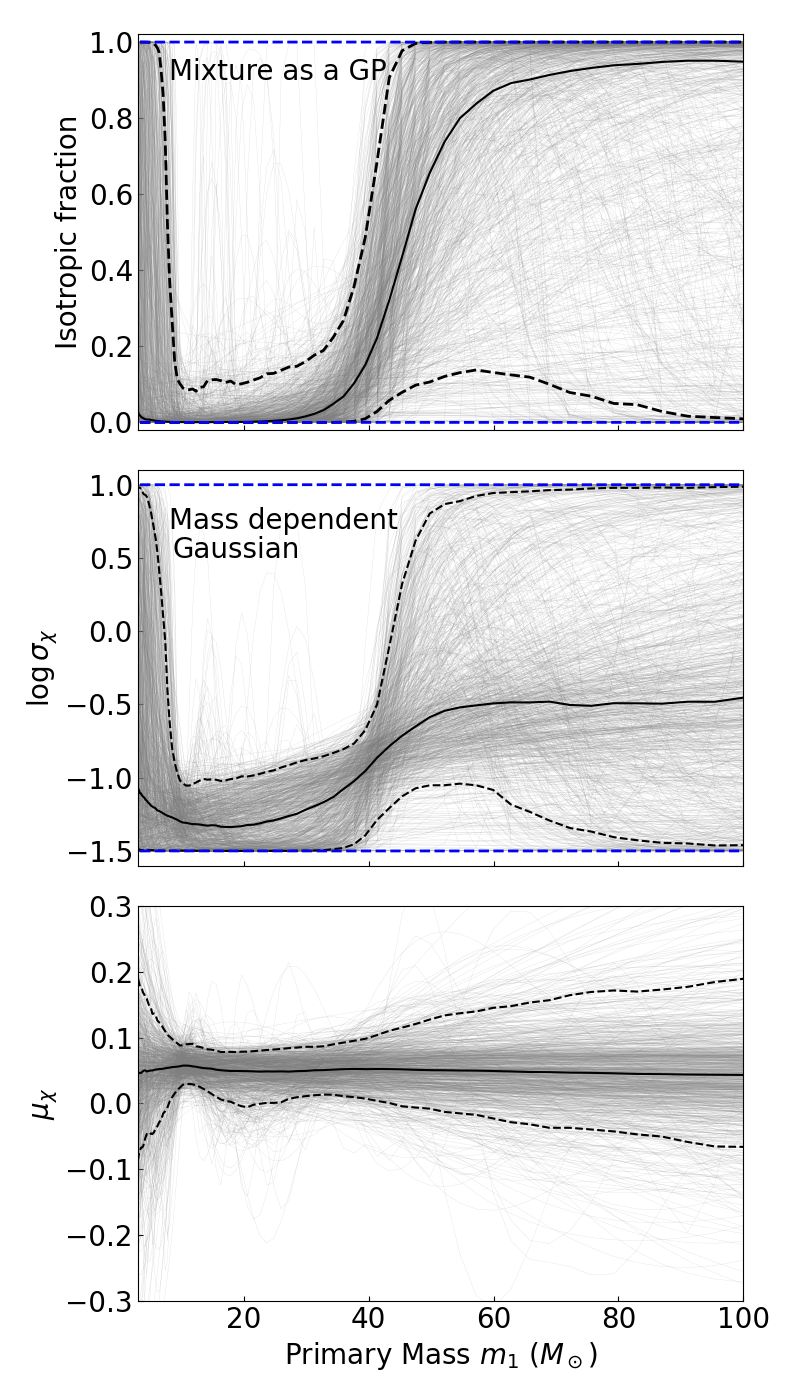}
\caption{
\textit{Upper panel:} 
Median and 90\% confidence intervals
of the mass-dependent mixture fraction \(\zeta(m_1)\) as a function of primary BH mass,  representing the relative contribution of the isotropically spinning BH population to the total merger rate. 
\textit{Middle and lower panels}:
the mean and standard deviation of the 
$\chi_{\rm eff}$ distribution, represented
here as a truncated Normal distribution with mass dependent mean and
variance. Blue-dashed lines show the prior 90\% confidence region. For $\mu_\chi$  these intervals are not shown as they are beyond the range of values plotted.
}
    \label{MFplots}
\end{figure}

\subsection{Mixture fraction}
\label{mix}
If a (P)PISNe mass gap exists, and it is repopulated by BHs formed from a previous merger, then above a certain mass threshold (i.e., the lower edge of the gap) the  
distribution of $ \chi_{\rm eff}$ should change and become consistent with equation~(\ref{CDFeff}).

In this section we will return to the model explored in Ref.~\cite{2025PhRvL.134a1401A}, in which the distribution of BBH spins is a mass-dependent mixture of two components: a truncated Gaussian and the broad and uniform distribution represented by equation~(\ref{CDFeff}).
Let $\zeta(m_1)$ be the fraction of events with uniformly-distributed spins,
such that 
\begin{equation}\label{eq:pi1}
\begin{aligned}
p(\chi_\mathrm{eff}|m_1) =(1 - \zeta(m_1))  \mathcal{N}(\chi_{\text{eff}};\mu,\sigma)+\\
\zeta(m_1)\mathcal{U}(\chi_\mathrm{eff};w=0.47)\ .
\end{aligned}
\end{equation}
Here, $\mathcal{N}(\chi_\mathrm{eff};\mu,\sigma)$ denotes a normalized Gaussian distribution with mean $\mu$ and standard deviation $\sigma$ truncated within $[-1, 1]$, and $\mathcal{U}(\chi_\mathrm{eff}\,;w)$ is a uniform distribution defined over the range $|\chi_\mathrm{eff}|<w$.
We set $w = 0.47$, as predicted for mergers involving 2nd generation BHs.

In Ref.~\cite{2025PhRvL.134a1401A}, we assumed that $\zeta(m_1)$ rose sharply from zero to one at the transition scale $m_1$, restricting our model to a single transition from a purely-Gaussian to a purely-uniform effective spin distribution.
Here, we instead model the mixture fraction $\zeta(m_1)$ via a GP over $\ln m_1$, allowing for considerably more complex behaviour in the mass-dependent spin distribution.
A GP model for $\zeta(m_1)$ allows for non-monotonic behaviour, such as multiple transitions back and forth between spin morphologies.
It moreover illustrates at which masses we have informative spin measurements, and thus well-measured $\zeta(m_1)$, and in which regions data are instead uninformative.
Finally, $\zeta(m_1)$ will allow us to bound the fraction of systems possibly arising from dynamical formation across the full range of BH masses.

We specifically model $\zeta(m_1)$ via
\begin{equation}\label{kzeta}
\zeta(m_1) = S\left(\Psi[\ln(m_1)]\right)
\end{equation}
where $\Psi(x)$ is drawn from a zero-mean GP prior, $\Psi(x) \sim \mathcal{GP}\left(0,\ k(x, x'; a_\zeta, \ell_\zeta)\right)$, using a squared exponential kernel with variance $a_{\zeta}^2$, and smoothing length
$\ell_{\zeta}$. 
A sigmoid function
\begin{equation}
S(x) = \frac{1}{1 + e^{-x}}
\end{equation}
is then applied to $\Psi(m_1)$ in order to ensure $0 \leq \zeta(m_1) \leq 1$.

The inferred \(\zeta(m_1)\)
is shown in the upper panel of Fig.~\ref{MFplots}.
It exhibits a clear trend with primary BH mass. At  \(m_1 \lesssim 5\,M_\odot\), \(\zeta(m_1)\) is unconstrained,  spanning the full range from 0 to 1. This indicates that the data are not informative in this regime: this is expected, given the low number of detections and the difficulty in distinguishing spin contributions at small masses. Between \(m_1 \approx 10\,M_\odot\) and \(40\,M_\odot\), the 95\% credible interval for \(\zeta(m_1)\) remains consistent with 0 and constrained below $\simeq 0.15$. This implies that in this mass range the contribution from a dynamically assembled, isotropically spinning population is statistically small, i.e., less than $15\%$ of the astrophysical population. The truncated Gaussian representing $\rm \chi_{\rm eff}$ is  a narrow distribution with positive mean $\mu=0.05^{+0.03}_{-0.03}$ ($90\%$ credibility).

Above \(m_1 \approx 40\,M_\odot\), we observe a marked increase in \(\zeta(m_1)\), with the mixture fraction bounded above 
\(\zeta(m_1)>0.1\)
 at $90\%$ credibility,
indicating a statistically significant non-zero contribution from a different population with a broader $\chi_{\rm eff}$ distribution. 
This result indicates a transition in the BH population around \(40\,M_\odot\). Above this mass, the lower $5\%$ bound of the isotropic spin fraction approaches the $95\%$  upper bound of the distribution below \(40\,M_\odot\). This indicates a significant shift in spin properties, consistent with the emergence of a distinct subpopulation at higher masses.

In Fig.~\ref{MFplots}, we  show the 90\% prior confidence region. The broad prior on the kernel parameter $a_{\zeta}$ leads to intervals that, after applying the sigmoid transformation, lie close to the prior boundaries. 
 The inferred 5\% lower bound on the isotropic spin fraction in the interval $\sim 40$ to $80~M_\odot$ remains significantly above zero. This indicates that the data are constraining this feature: the rise in $\zeta(m_1)$ at high masses is not simply reflecting the edge of a diffuse prior or of an informative region, but rather represents a data-driven trend supported by the observed  population.
At higher masses still, data again become uninformative due to a low number of observations, and $\zeta(m_1)$ asymptotes back to its prior.

\subsection{Effective spin $vs$ primary mass }\label{musigma}
To further investigate possible correlations between spin and primary mass, we consider a model in which the effective spin parameter \(\chi_{\mathrm{eff}}\) is instead described as a single Gaussian, but one whose mean and standard deviation vary as functions of the primary BH mass \(m_1\):
\begin{equation}\label{mu_sig_m}
p (\chi_{\mathrm{eff}}|m_1) = \mathcal{N}(\mu_{\chi}(m_1),\, \sigma_{\chi}(m_1))
. \end{equation}
Here, $\mathcal{N}(\chi_{\mathrm{eff}}; \mu, \sigma)$ denotes a truncated normal distribution
where both the mean \(\mu_{\chi}(m_1)\) and the standard deviation  \( \sigma_{\chi}(m_1)\) are modeled as non-parametric functions using independent GPs over \(\log(m_1)\) and adopting an exponential kernel with parameters $a_\mu$, $\ell_\mu$, $a_\sigma$ and $\ell_\sigma$. 
The prior range of \( a_\mu \) is then constrained to the interval \([-1, 1]\) using  the sigmoid function $2S(x)-1$, while \( \log \sigma_{\chi} \)  is constrained to the interval \([-1.5, 1]\) using $2.5S(x)-1.5$. 
The mass distribution is again represented non-parametrically as described in Sec.~\ref{sec:methods}.

In contrast to mixture models that impose a population split (e.g., below and above a threshold mass), this framework enables us to search for continuous or gradual changes in the spin distribution that may signal a gradual shift in the population, rather than a sharp transition to a different population above a certain mass (as in Ref.~\cite{2025PhRvL.134a1401A}) or an evolving mixture between two starkly different spin distributions (as in Sec.~\ref{mix}). A sharp increase in the variance of \(\chi_{\mathrm{eff}}\) or a change in its mean above a certain value of \(m_1\) would be consistent with the emergence of a different population. 

The results of this analysis are presented in the middle and lower panels of Fig.~\ref{MFplots}. 
The mean of the \(\chi_{\rm eff}\) distribution is  measured away from zero at $m_1 \simeq 10~M_\odot$, where the merger rate has a strong peak (see Fig.~\ref{fig:massplot}). This suggests that the majority of merging binaries at these masses are unlikely to originate solely from dynamical interactions in dense star clusters. A positive mean for the \(\chi_{\rm eff}\)
  distribution indicates the presence of a subpopulation with preferentially aligned spins, consistent with formation channels  such as isolated binary evolution in the field \cite[e.g.,][]{2023arXiv230401288G,Mandel2016a}, 
or formation in AGN disks \cite[e.g.,][]{Bartos2016}.

Across the entire mass range, the mean of the \(\chi_{\rm eff}\) distribution shows little variation with primary mass, although the uncertainty increases with mass such that the mean becomes consistent with 
zero above $35M_\odot$ ($90\%$ credibility).
The standard deviation exhibits clear mass dependence, however, closely mirroring the behaviour of \(\zeta(m_1)\). 
Between \(5\) and \(40\,M_\odot\), the distribution is well described by a narrow Gaussian with a positive mean of approximately \(0.05\) and a standard deviation corresponding to \(\log \sigma \simeq -1.3\).
At masses above \(40\,M_\odot\), the dispersion increases and becomes  \(\log \sigma \gtrsim -1\) ($90\%$ credibility). 
These results are consistent with two distinct populations below and above \(\simeq 45\,M_\odot\), that are characterised by different spin distributions. We note that a similar result was recently obtained by Ref.~\cite{2025arXiv250602250S}, who inferred  the joint component mass and  $\chi_{\rm eff}$ distribution
   using an iterative kernel density estimation-based method.

In Ref.~\cite{2025PhRvL.134a1401A}, we showed that the low-mass population is well characterized by a narrow Gaussian, while the high-mass population exhibits a broader distribution.
We argued that this high-mass component is responsible for the trend observed in previous studies, where the spin dispersion was found to increase with mass under parametric models assuming a linear dependence of the mean and variance on primary mass \cite{2022ApJ...932L..19B}.
Consistently, the present analysis shows that such linear models are unlikely to be a good representation of the real trend \cite[see also][]{2024PhRvD.109j3006H}.
However, our analysis does not fully rule out a smoother transition or gradual trend in the spin properties of the overall population.
For instance, as shown in Fig.~\ref{MFplots}, several posterior sample tracers exhibit a more continuous or approximately linear increase of $\log \sigma$ with mass.

\begin{figure} 
\includegraphics[width=\columnwidth]{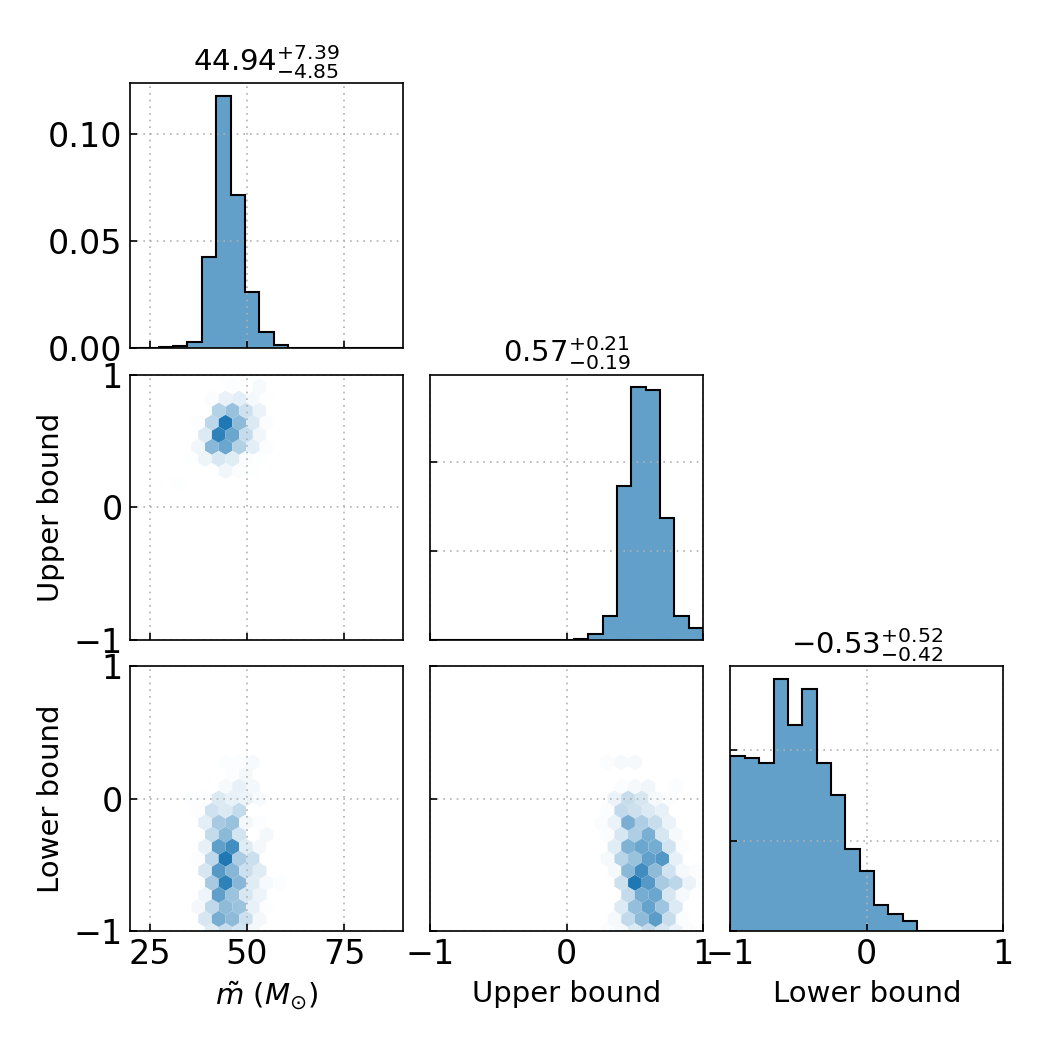} 
    \caption{The lower and upper bounds of the $\chi_{\mathrm{eff}}$ distribution inferred using the parametric model of equation~(\ref{UL_ind}), as well as the posterior of the transition mass $\tilde{m}$ between the low mass and high-mass/high-spin BH populations.
   }
    \label{fig:IndBounds}
\end{figure}

\begin{figure} 
\includegraphics[width=\columnwidth]{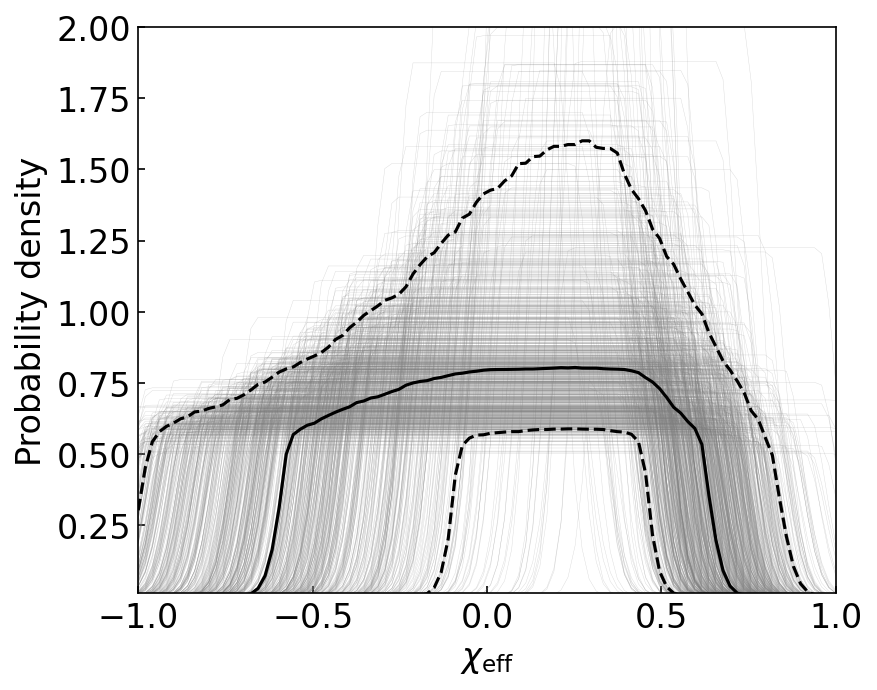} 
    \caption{
The $\chi_{\mathrm{eff}}$ distribution inferred using the parametric model of equation~(\ref{UL_ind}), where the high mass population is represented by a uniform distribution with independent lower and upper bounds. The solid line is the posterior median, while the dashed lines show the $90\%$ confidence intervals.
 Individual posterior draws  are shown via light gray
traces.
    }
    \label{fig:IndBounds2}
\end{figure}

\begin{figure*} 
\includegraphics[width=0.9\textwidth]{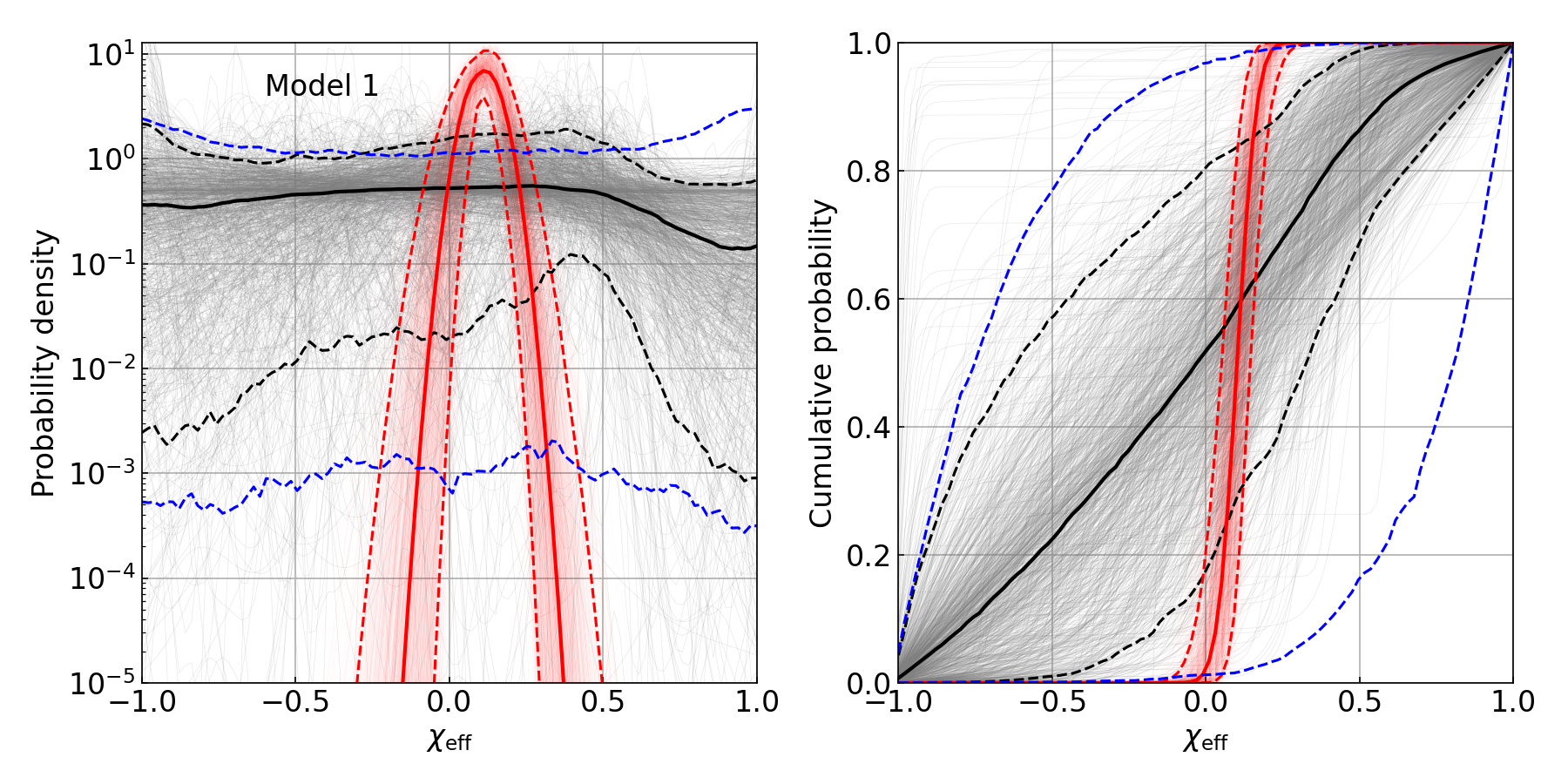} 
\includegraphics[width=0.9\textwidth]{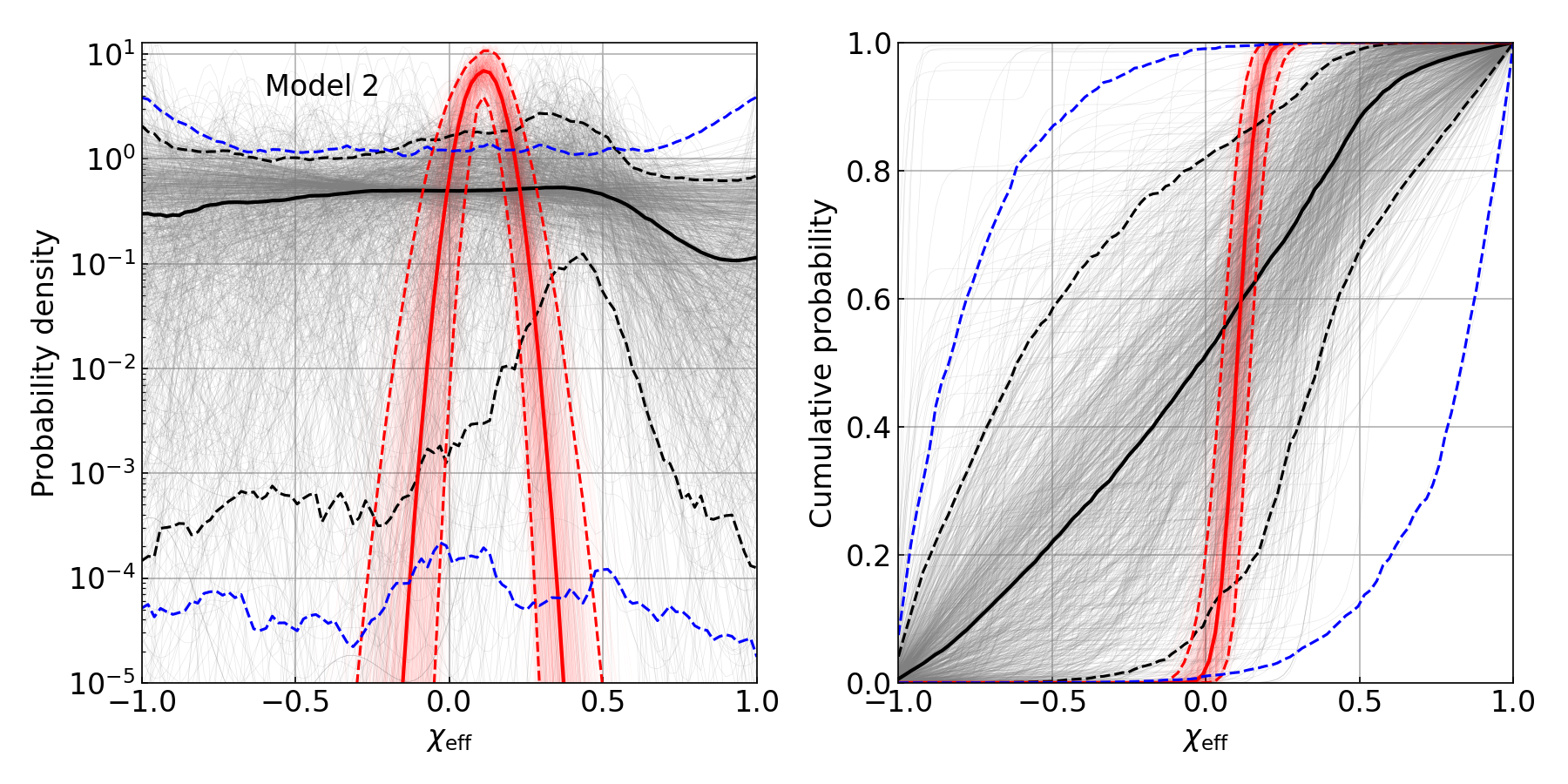} 
    \caption{The distribution of $\chi_{\rm eff}$ obtained under model~\eqref{Xeff_GP} for both the Gaussian component at $m_1< \tilde{m}$ (red)
and the non-parametric component at 
$m_1\geq \tilde{m}$ (black and green). We show the results for
two different choice of priors on the kernel parameters used in  hierarchical analysis, showing how these affect the resulting distributions.
   Individual posterior draws are shown via light gray and red traces. 
 Blue-dashed lines show the prior $90\%$ confidence intervals.}
    \label{fig:XeffD}
\end{figure*}

\begin{figure} 
\includegraphics[width=\columnwidth]{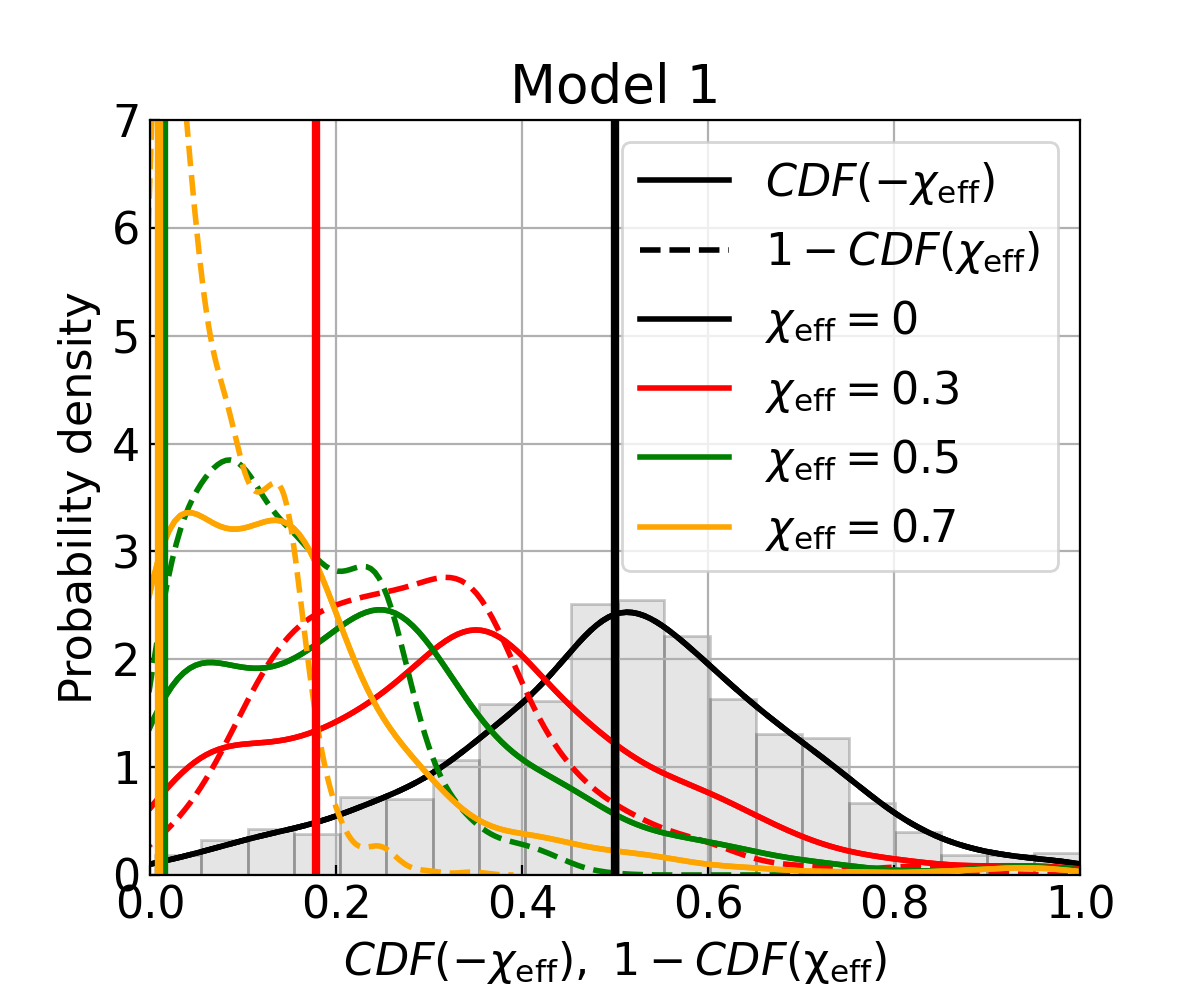}\\ \includegraphics[width=\columnwidth]{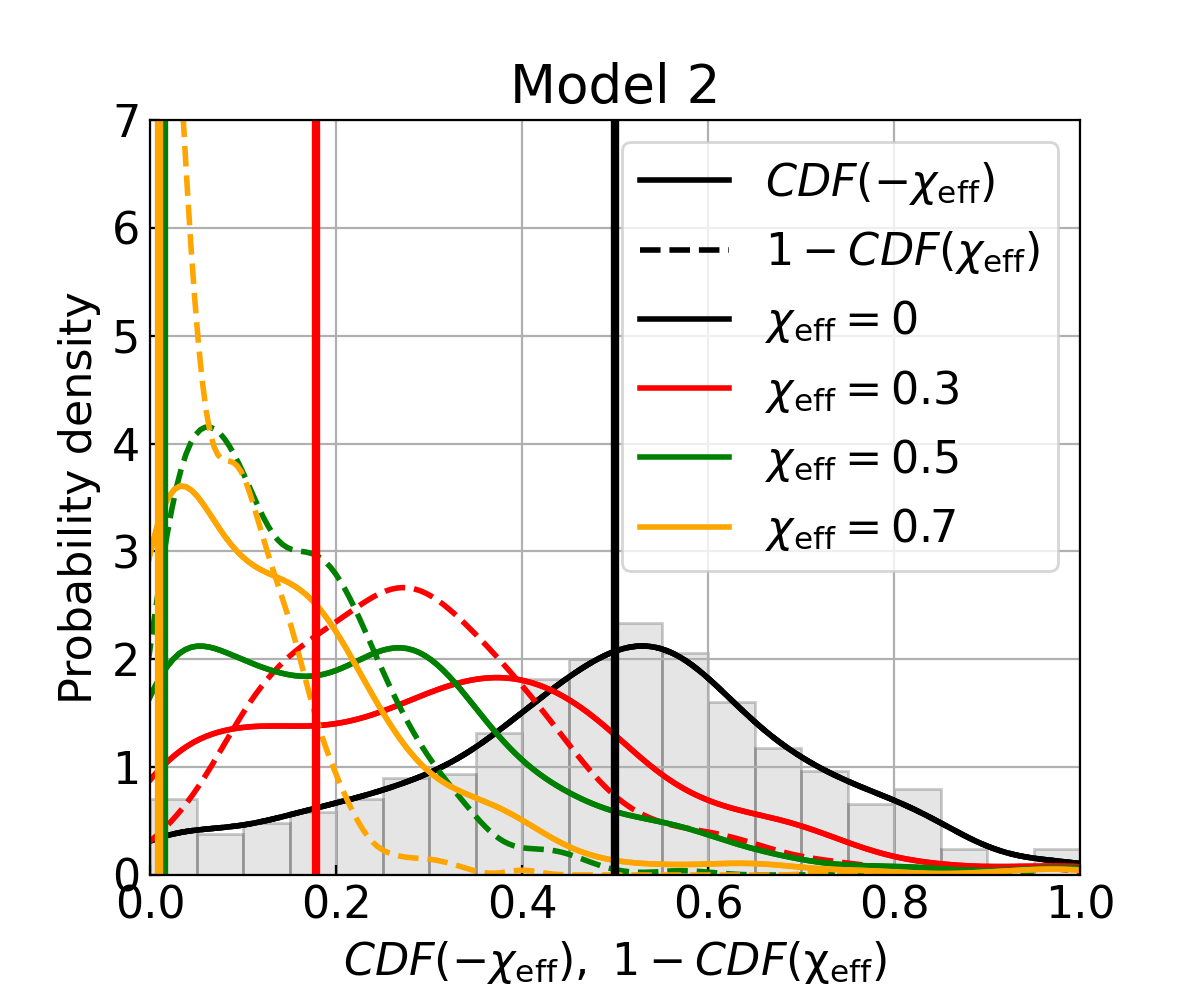} 
    \caption{
    Posterior distributions of the $\mathrm{CDF}$ of  $\chi_{\rm eff}$ evaluated at selected values: $\chi_{\rm eff} = 0,\ 0.3,\ 0.5,$ and $0.7$. We  show the posteriors for $\mathrm{CDF}(-\chi_{\rm eff})$ and $1 - \mathrm{CDF}(\chi_{\rm eff})$ to assess the symmetry of the $\chi_{\rm eff}$ distribution about zero. For an isotropic distribution, we expect $\mathrm{CDF}(\chi_{\rm eff}=0) = 0.5$ and $\mathrm{CDF}(-\chi_{\rm eff}) = 1 - \mathrm{CDF}(\chi_{\rm eff})$ at each value. The vertical lines give the values expected under the dynamical formation hypothesis, derived from equation~(\ref{CDFeff}). In grey we show the unsmoothed histogram giving the distribution of the cumulative function evaluated at \(\chi_{\rm eff} = 0\), i.e., \( {\mathrm{CDF}}(\chi_{\rm eff} = 0) \). The two panels correspond to different prior choices for the kernel parameters used in the GP as explained in the main text.
    }
    \label{fig:simmetry}
\end{figure}

\subsection{Symmetry of the effective spin distribution}

As identified in Ref.~\cite{2025PhRvL.134a1401A} and corroborated above with more flexible models, current data strongly support a correlation between BH spins and mass, with BHs above masses $\tilde m \approx 45\,M_\odot$ exhibiting a broader spin distribution than those below.
In Ref.~\cite{2025PhRvL.134a1401A}, we deliberately enforced the spin distribution of  high-mass binary BHs to be isotropic (i.e. an effective spin distribution that is uniform and symmetric about zero).
This choice was motivated by the possibility of second-generation BH mergers in dense environments, which robustly produce isotropic spins.

Here, we seek to more carefully test this hypothesis.
In particular, do the effective spins of massive binary BHs indeed extend to both larger positive and larger negative values than their less massive counterparts?
Or are conclusions about large and negative effective spins primarily driven by prior choices and modeling assumptions?

\subsubsection{Parametric: Independent minimum and maximum bounds}

We first proceed parametrically.
As in Ref.~\cite{2025PhRvL.134a1401A}, we adopt an effective spin model that transitions from a Gaussian to a uniform distribution below and above $\tilde m$, respectively.
We extend this model, however, by regarding the upper ($\chi_{\rm eff, max}$) and lower truncation bounds ($\chi_{\rm eff, min}$) of the uniform distribution as independent parameters to be inferred from data:
    \begin{equation}
    \label{UL_ind}
    p_{\mathcal{N}+\mathcal{U}}(\chi_{\mathrm{eff}}\,|\,m_1) =
    \begin{cases}
    \mathcal{N}(\chi_{\mathrm{eff}};\, \mu, \sigma), & m_1 < \tilde{m} \\
    \mathcal{U}(\chi_{\mathrm{eff}};\, \chi_{\rm eff, min}, \chi_{\rm eff, max}) & m_1 \geq \tilde{m} .
    \end{cases}
    \end{equation}
Here, $\mathcal{U}(\chi_{\mathrm{eff}}; \chi_{\rm eff, min}, \chi_{\rm eff, max})$ denotes a uniform distribution over the interval $[\chi_{\rm eff, min}, \chi_{\rm eff, max}]$.
We place uniform priors on the lower and upper bounds of the uniform component: $ \chi_{\rm eff, max}\in \mathcal{U}(0.05, 1)$ and  $\chi_{\rm eff, min} \in \mathcal{U}(-1,\chi_{\rm eff, max})$. 

The posterior  of the parameters governing the $\chi_{\mathrm{eff}}$ distribution for the high-mass BH population are shown in Fig.~\ref{fig:IndBounds}. 
The posterior distribution of $\chi_{\mathrm{eff}}$  is shown in Fig.~\ref{fig:IndBounds2}.
The inferred transition mass is found to be $\tilde{m} = 45^{+7}_{-5}\,M_\odot$, and the median of the $\chi_{\mathrm{eff}}$ distribution is
inferred to be $-0.01^{-0.21}_{-0.24}$ (90\% credibility).
The upper bound of the uniform distribution is relatively well constrained by the data, with $\chi_{\rm eff, max} = 0.57^{+0.21}_{-0.19}$.
The lower bound $\chi_{\rm eff, min}$, however, is more weakly constrained.
We find that $\chi_{\rm eff, min} < 0$ at $98\%$ credibility, indicating  evidence for negative effective spin among high-mass binary BHs.
At the same time, we cannot confidently exclude a minimum effective spin that is zero or small and positive.
We therefore conclude that the broadening effective spin distribution at high masses is primarily driven by massive BHs exhibiting larger, more positive spins.
Whether massive BHs also exhibit larger and more negative spins remains unclear with current data.

Despite the weak constraints on $\chi_{\rm eff, min}$, we note that current data remain consistent with expectations from repeated mergers in dense clusters.
In a dynamical formation scenario where the primary BH formed from a previous merger we should expect $\chi_{\rm eff, max}=-\chi_{\rm eff, min}\simeq 0.5$ \cite{2025PhRvL.134a1401A}.
The results of Fig.~\ref{fig:IndBounds} and Fig.~\ref{fig:IndBounds2} illustrate that the data are compatible with this basic prediction; additional data will be required to further confirm or rule out this hypothesis.

\subsubsection{Non-parametric: Gaussian process effective spin distributions}\label{npm}

As a further check on these conclusions, we consider a more flexible version of Eq.~\eqref{UL_ind} in which the spin distribution at high masses is described via a GP prior:
    \begin{equation}
    \label{Xeff_GP}
    p(\chi_{\mathrm{eff}}\,|\,m_1) =
    \begin{cases}
    \mathcal{N}(\chi_{\mathrm{eff}};\, \mu, \sigma) & m_1 < \tilde{m} \\
    {e^{\Theta(\chi_{\mathrm{eff}})}}/{\int_{-1}^{1} e^{\Theta(\chi_{\rm eff})}\, d\chi_{\rm eff}}& m_1 \geq \tilde{m} \ .
    \end{cases}
    \end{equation}
Here, the function $\Theta(\chi_{\mathrm{eff}})$ is generated from a GP,
    \[
    \Theta(\chi_{\mathrm{eff}}) \sim \mathcal{GP}\left(0,\, k(\chi_{\mathrm{eff}}, \chi_{\mathrm{eff}}'; \ a_\chi, \ell_\chi)\right),
    \]
with zero mean and a squared-exponential covariance kernel. 
We evaluate these GPs on a regular grid of $N_{\rm bin}=100$ points in $\chi_{\mathrm{eff}}$ within the range $-1$ to $+1$.
We also consider an additional  fit
where the non-parametric model for the primary mass distribution was replaced by the  power law+Gaussian peak  model often used in the literature \cite[e.g.,][]{2023PhRvX..13a1048A}.  The results, however,
remain nearly unchanged and  are not reported here. This shows, however, that the inference about spin properties is robust across different assumptions about the BH mass spectrum.


We show the inferred distribution of $\chi_{\mathrm{eff}}$  in  Fig.~\ref{fig:XeffD}. The low and high mass populations can be clearly separated based on their $\chi_{\mathrm{eff}}$ distributions.
We find that the lower bound of the $\chi_{\rm eff}$ distribution for the high mass population  depends somewhat on the prior choice for the kernel parameters. Thus, in Fig.~\ref{fig:XeffD}
we show the results for  two different choices of prior that are given in
 Table~\ref{tab:priors}
 of Appendix~\ref{priors}. As shown in the figure,   Model 2 allows for larger amplitude variations of $e^{\Theta(\chi_{\mathrm{eff}})}$ over the  $\chi_{\rm eff}$ grid than Model 1, resulting in a broader prior.

The two prior models produce overall similar distributions for \(\chi_{\rm eff}\). 
Specifically, the inferred median of
the $\chi_{\rm eff}$ distribution under the two models  are
$-0.03^{+0.36}_{-0.59}$ (Model 1) and
$-0.02^{+0.34}_{-0.59}$ (Model 2).
However, we observe noticeable differences in the lower $5\%$ credible bound outside the range \(0 \lesssim \chi_{\rm eff} \lesssim 0.7\), where Model 2 leads to significantly lower probability values. This indicates that the current gravitational-wave data provide only weak constraints on the shape of the \(\chi_{\rm eff}\) distribution in these regions. Consequently, the inferred distribution in this range is more sensitive to the choice of prior, reflecting the limited information content of the data.

An interesting feature of the $\chi_{\rm eff}$ posterior is the asymmetry observed in the lower tail of the distribution, which tends to extend more prominently toward positive values. This asymmetry has previously been interpreted as a physical signature in the $\chi_{\rm eff}$ distribution, possibly pointing to formation channels involving AGN disks, where accretion disks can efficiently align BH spins with the orbital angular momentum \cite{2025arXiv250109495L}. Such alignment mechanisms would naturally lead to a positive skew in $\chi_{\rm eff}$ due to preferentially aligned spins. 
Alternatively, it could still be possible
that the  (P)PISN gap stars at much larger mass values
and that the positive alignment is
produced by binary stellar evolution where tides
can align the stellar progenitor spins with the orbital angular momentum \cite{Hut1981}.
However, our interpretation is that the observed asymmetry is not necessarily physical, but rather a consequence of the data-limited ability to constrain the negative end of the $\chi_{\rm eff}$ distribution. This is particularly evident when examining the cumulative distributions, which reveal that the lower edge of the posterior spans a much broader range than the upper edge. This broad lower tail reflects the fact that large negative values of $\chi_{\rm eff}$ remain poorly constrained, leading to an apparent-but not statistically significant-asymmetry.

In Fig.~\ref{fig:simmetry} we show the posterior distributions of the $\mathrm{CDF}$ evaluated at specific values of $\chi_{\rm eff}$. In particular, we display the posteriors for $\mathrm{CDF}(\chi_{\rm eff}=0)$, as well as for $\mathrm{CDF}(-\chi_{\rm eff})$ and $1 - \mathrm{CDF}(\chi_{\rm eff})$ evaluated at $\chi_{\rm eff} = 0.3,\ 0.5,$ and $0.7$. For a population with isotropic spin orientations and randomly aligned angular momenta, the $\chi_{\rm eff}$ distribution is expected to be symmetric about zero. This symmetry implies the two conditions: (1) a median $ \mathrm{CDF}(\chi_{\rm eff}=0) = 0.5$, and (2) $\mathrm{CDF}(-\chi_{\rm eff}) = 1 - \mathrm{CDF}(\chi_{\rm eff})$ at a given $\chi_{\rm eff}$.

From the posterior distributions shown in Fig.~\ref{fig:simmetry}, we find that both symmetry conditions (1) and (2) are satisfied within the uncertainties of the data. Specifically, the median values of the CDFs are consistent with the expectations for an isotropic spin distribution across the selected values of $\chi_{\rm eff}$. However, the posterior distributions remain relatively broad, highlighting the limited constraining power of current observations. In particular, the posterior for $\mathrm{CDF}(\chi_{\rm eff}=0)$ retains non-zero support at zero, indicating that highly asymmetric spin distributions—such as those strongly peaked at positive $\chi_{\rm eff}$ values—are disfavored but not definitively excluded by the data. Conversely, the data do rule out a complete absence of systems with positive $\chi_{\rm eff}$, implying that a subpopulation of spin-aligned binaries is present in the high mass population.

\section{Astrophysical Implications}
\label{sec:astro}
The mass transition we identified in this work  has important implications for the formation and evolution of BBHs, and for the astrophysical environments in which they form. The emergence of a broader \(\chi_{\rm eff}\) distribution above \(\simeq 45\,M_\odot\), observed consistently across multiple models, points to the  presence of a distinct high-mass subpopulation. This transition mass lies near the expected onset of the (P)PISN mass gap \cite{2019ApJ...887...72L,2020ApJ...902L..36F,2023ApJ...950L...9S} and suggests that BHs above this threshold are not formed through ordinary stellar collapse.

\begin{figure}
    \centering \includegraphics[width=1\columnwidth]{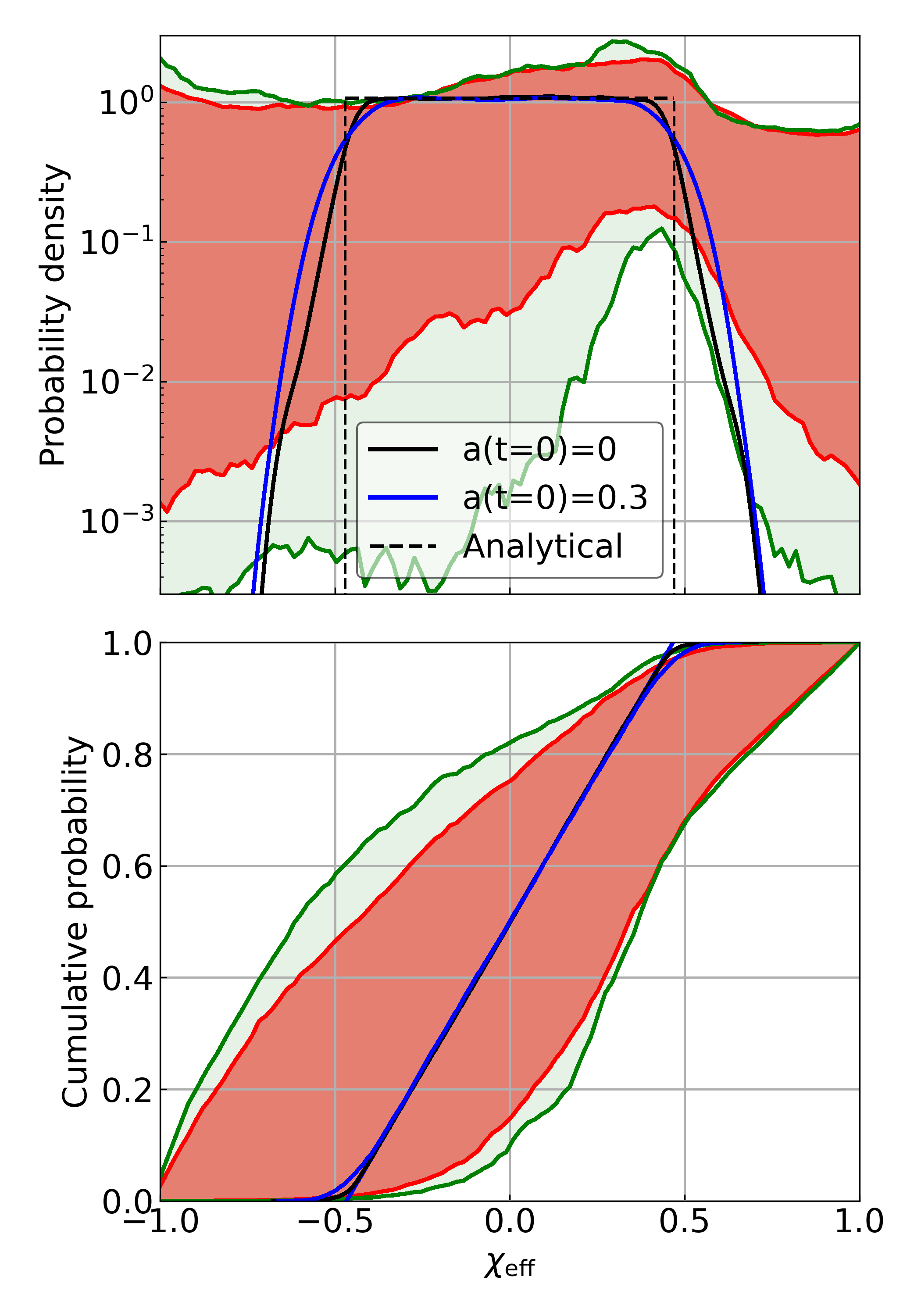}
    \caption{
    Comparison between the inferred $\chi_{\mathrm{eff}}$ distribution and the distribution predicted by cluster models within the pair-instability mass gap. We show the $90\%$ confidence intervals inferred using equation~\eqref{Xeff_GP} and for our two prior choices: Model 1 (red) and Model 2 (green). The blue dotted line is the simple analytical model of equation~\eqref{CDFeff}. The continuous black and blue lines are obtained from cluster models that were evolved with \texttt{cBHBd} \citep{2020MNRAS.492.2936A}. We show the results of two different choices for the natal spins of the BHs, in one all BHs are initially non-spinning, and in the other model the all have an initial dimensionless spin parameter $a=0.3$.
    }\label{comp}
\end{figure}

One possible origin of the high mass population we identified is dynamical formation in a dense stellar environment where the primary BH was formed from a previous merger.
Such environments include globular clusters \cite[e.g.,][]{2017ApJ...834...68C,2018PhRvL.120o1101R}, young massive clusters \cite[e.g.,][]{2019MNRAS.483.1233R,2021MNRAS.507.5132D,2021MNRAS.503.3371B}, and nuclear star clusters \cite[e.g.,][]{2016ApJ...831..187A,2019MNRAS.486.5008A}, where frequent dynamical interactions can assemble and harden BH binaries, leading to repeated mergers over time. To assess whether this hypothesis is consistent with current data, we evolve a grid of star cluster models using the code \texttt{cBHBd} \citep{2020MNRAS.492.2936A}. Since the $\chi_{\rm eff}$ distribution is largely insensitive to the choice of initial cluster conditions \citep{2025PhRvL.134a1401A}, we simulate a single population of 10{,}000 clusters with fixed metallicity $Z = 1\%\,Z_{\odot}$ and with masses uniformly distributed in logarithmic space between $10^3$ and $10^6\,M_{\odot}$. The clusters are evolved for 10\,Gyr. In Fig.~\ref{comp}, we show the resulting $\chi_{\rm eff}$ distribution for all mergers occurring in the pair-instability mass gap, which in our models lies at $\simeq 50\,M_{\odot}$. We find that the predicted distributions lie almost entirely within the 90\% credible intervals of the distribution inferred from the data. Small differences in the tails of the distributions are not considered significant, as those regions are partially shaped by prior assumptions and are not strongly constrained by the data.

If a significant fraction of the high-mass BBH mergers are indeed dynamically assembled—as our analysis tentatively indicates—this would reinforce the idea that dense stellar clusters play a central role in populating the upper end of the BH mass distribution. 
Dynamical formation models in which clusters dominate the high-mass BH merger rate often also predict a significant, if not dominant, contribution from cluster-formed binaries at lower masses, below the expected pair-instability gap threshold \cite{2023MNRAS.522..466A}. Thus, identifying a dynamically assembled subpopulation not only sheds light on the high-mass regime but will also put constraints on the contribution of clusters across the full BH mass spectrum. More broadly, evidence for a cluster origin provides a valuable probe of internal cluster dynamics—offering constraints on BH retention efficiency following gravitational recoil, the role of mass segregation in binary formation, and the initial structural and dynamical conditions required for clusters to retain and recycle merger remnants.

On the other hand, if future data were to rule out a pure cluster origin for the high-mass, broad-spin population, this would also carry significant astrophysical consequences. This would imply that the apparent lack of a sharp pair-instability mass gap in the observed population is not the result of mass growth through BH mergers. In this case, the gap may occur at significantly higher masses than predicted, or might be absent altogether, calling into question existing models of stellar evolution, particularly the treatment of (P)PISN. Alternatively, other formation channels could be contributing to the high-mass population. These include mergers in AGN disks, where gas torques and accretion can aid binary formation and potentially align spins \cite{2021ApJ...908..194T,2024MNRAS.531.3479M}, or consecutive mergers occurring in the field through the evolution of triples and higher-order systems \cite{2019MNRAS.486.4781F,2020ApJ...895L..15F}. 
Another  scenario that can produce high-spin, aligned binaries is chemically homogeneous evolution  \citep[e.g.,][]{Marchant2016, Mandel2016,Fuller2019, Riley2021,DallAmico2025}. In systems with orbital periods $\lesssim1\text{–}2\,$days, strong tidal forces spin the stars up to near their critical velocities, driving efficient Eddington–Sweet circulations and shear instabilities that mix hydrogen from the envelope into the core and helium outward.  This rotational mixing prevents the development of a deep convective envelope and suppresses radial expansion on the main sequence, so the components remain compact and tidally locked throughout their lives.  Consequently, at core collapse, each star retains most of its angular momentum, producing BHs with near-maximal spins that are almost perfectly aligned with the orbital angular momentum.  The resulting $\chi_{\rm eff}$ distribution is then sharply peaked at high positive values, in stark contrast to the broad, symmetric distribution expected from dynamical channels.   
Such scenarios typically make different predictions for spin and eccentricity distributions, redshift evolution, and merger rates, and distinguishing between them will be a major goal for future observational campaigns.

Identifying the location of the inferred transition mass will  provide indirect constraints on stellar evolution and nuclear astrophysics. The pair-instability mass gap depends sensitively on the properties of massive stars, particularly on the helium core mass at collapse, which is set by the rate of the \( {}^{12}\mathrm{C}(\alpha,\gamma){}^{16}\mathrm{O} \) nuclear reaction during helium burning \cite{Woosley2016}. A transition to second-generation mergers at \(\simeq 45\,M_\odot\) would be consistent with the presence of a (P)PISN mass gap at the lower end of the final BH masses than is sometimes assumed. If confirmed, this could favour models with relatively high \( {}^{12}\mathrm{C}(\alpha,\gamma){}^{16}\mathrm{O} \) rates, leading to more massive CO cores and earlier onset of instability. While current data do not yet allow strong constraints on this reaction rate, future improvements in mass and spin measurements could strengthen the connection between gravitational-wave astronomy and nuclear physics.

 Our results also have potential implications for standard siren cosmology \cite{2025arXiv250210780T}.  If the sharp transition from low to high effective spin values is indeed associated with the onset of 1st+2nd generation mergers, it would constitute a relatively redshift-invariant mass scale in the source frame. As such, it provides a physically motivated and stable reference point for spectral siren cosmology, complementary but more robust than, for example, the Gaussian excess at \( \sim 35\,M_\odot \), whose astrophysical origin remains uncertain.
However, we note that if metallicity evolution causes the mass gap location to drift with redshift, employing the gap as a standard mass scale for spectral sirens cosmology could introduce systematic offsets in distance estimates.  As a result, improving nuclear reaction rate measurements and mass loss models is vital not only to map BH formation channels, but also to establish the (P)PISN gap as a robust tool in both astrophysical population studies and precision cosmology \citep{2016A&A...594A..97B,Farmer2020}.

It is also important to recognize that theoretical uncertainties in the precise location and width of the (P)PISN mass gap - driven by factors such as the $^{12}\mathrm{C}(\alpha,\gamma)^{16}\mathrm{O}$ reaction rate \citep{Shen2023}, wind mass loss prescriptions, progenitor rotation, and internal mixing \citep{Marchant2020} - can shift both the lower and upper edges of the gap by tens of solar masses \citep{Woosley2021}.  Such shifts will naturally smooth out the predicted black-hole mass function, turning what might appear as a sharp threshold in $\chi_{\rm eff}$ into a more gradual transition.  In practice, this means that the high spin population above $\simeq 45\,M_\odot$ could partially arise from the smearing of the gap itself, rather than solely from a distinct second-generation subpopulation; disentangling these effects will require tighter constraints on stellar evolution physics and more extensive gravitational wave catalogues \citep{Woosley2016,Marchant2019}.


Taken together, our results illustrate the potential of gravitational-wave observations not only to inform BH formation scenarios but also to probe stellar interiors and the dynamical environments in which massive binaries form. Continued accumulation of high-mass merger detections—particularly those with non-zero spin measurements—will be crucial for refining constraints on the properties of the high-mass BH population and clarifying its astrophysical origin.


\section{Summary and Conclusions}
\label{sec:conclusions}
 We have investigated the effective spin distribution of BBH mergers using non-parametric models that allow for  flexibility in both the spin and mass distributions. We find that above a primary BH mass of $ 46^{+7}_{-6}M_\odot$ the spin distribution 
becomes broader and more consistent with isotropic orientation of the spins and orbit. This value of the transitional mass is in good agreement with our previous work based on  strongly parametric models where the high-mass and isotropically spinning population was modeled using a uniform distribution, and where we inferred  
\(\tilde{m} = 44^{+6}_{-4}\,M_\odot\)
\cite{2025PhRvL.134a1401A}.

We explore the phenomenology of this mass-spin correlation using a variety of increasingly flexible models.
When adopting a Gaussian process model for the mass-dependent fraction $\zeta(m_1)$ of binary BH mergers with isotropic spins, we find that this fraction rises sharply above \(m_1 \approx 40\,M_\odot\), reaching \(\zeta(m_1) > 0.1\) at 90\% confidence and continuing to increase at higher masses.
Similar behavior is seen when instead modeling effective spins as a Gaussian distribution with a mass-dependent mean $\mu(m_1)$ and standard deviation $\sigma(m_1)$.
In the low-mass regime (\(m_1 < 40\,M_\odot\)), we  find the effective spin distribution to be tightly clustered around \(\mu_{\chi} = 0.05^{+0.03}_{-0.03}\) with a standard deviation \(\log \sigma \simeq -1.3\), consistent with low natal spins and spin alignment;
at higher masses, the width of the \(\chi_{\rm eff}\) distribution increases, with \(\log \sigma > -1\) at 90\% confidence, indicative of a broader spin distribution.

We furthermore explored the evidence for spin isotropy (i.e. symmetry in the effective spin distribution) among high-mass BHs.
When independently measuring the maximum and minimum effective spin among high-mass, we constrain the upper bound to \(\chi_{\rm eff, max} = 0.57^{+0.21}_{-0.19}\), while the lower bound remains weakly constrained at \(\chi_{\rm eff, min} = -0.53^{+0.52}_{-0.42}\).
However, 98\% of the posterior mass for \(\chi_{\rm eff, min}\) lies below zero, providing weak evidence for negative effective spins in this subpopulation.
Under our most flexible non-parametric models, the effective spin distribution  for the high-mass BBH population is consistent with being symmetric about zero, although current constraints remain limited by statistical uncertainties. 

The data do not exclude alternative possibilities, such as a spin distribution with an excess of aligned systems.
In particular, weak constraints on $\chi_{\rm eff, min}$ mean that we cannot rule a minimum effective spin among high-mass systems that is zero or small and positive.
In contrast, the upper bound is better constrained, with a posterior consistent with \(\chi_{\rm eff, max} \simeq 0.5\), as predicted for 1st+2nd generation mergers in a dense cluster.
This asymmetry in constraint quality—tighter on the upper end and looser on the lower—can give rise to an apparent skew in the posterior distribution of \(\chi_{\rm eff}\).
However, this should not be taken as  evidence for an intrinsically asymmetric spin population. Rather, it likely reflects the limited number of current gravitational-wave observations.
Given these limitations, we cannot make a definitive statement about the symmetry of the underlying distribution a part that  the data remain  consistent with a symmetric, isotropic spin distribution.
Putting more stringent constraints on a dynamical formation scenario will require a larger sample of high-mass mergers from current and future observing runs.

\begin{acknowledgments}
The authors would like to thank Jack Heinzel, and Matthew Mould for reviewing an early version of this manuscript and for their useful comments that helped to improve the paper.

FA and FD are supported by the UK’s Science and Technology Facilities Council grant
ST/V005618/1. IMRS acknowledges the support of the Herchel Smith Postdoctoral Fellowship Fund and the Science and Technology Facilities Council grant number ST/Y001990/1. This material is based upon work supported by
NSF’s LIGO Laboratory which is a major facility fully funded by
the National Science Foundation, as well as the Science and Technology Facilities Council (STFC) of the United Kingdom, the Max-Planck-Society (MPS), and the State of Niedersachsen/Germany for
support of the construction of Advanced LIGO and construction and
operation of the GEO600 detector. Additional support for Advanced
LIGO was provided by the Australian Research Council. Virgo is
funded, through the European Gravitational Observatory (EGO), by
the French Centre National de Recherche Scientifique (CNRS), the
Italian Istituto Nazionale di Fisica Nucleare (INFN) and the Dutch
Nikhef, with contributions by institutions from Belgium, Germany,
Greece, Hungary, Ireland, Japan, Monaco, Poland, Portugal, Spain.
KAGRA is supported by Ministry of Education, Culture, Sports, Science and Technology (MEXT), Japan Society for the Promotion of
Science (JSPS) in Japan; National Research Foundation (NRF) and
Ministry of Science and ICT (MSIT) in Korea; Academia Sinica
(AS) and National Science and Technology Council (NSTC) in Taiwan. The authors are grateful for computational resources provided
by Cardiff University and supported by STFC grant
ST/V005618/1. D.C. thanks the Gordon and Betty Moore Foundation for funding this research through Grant GBMF12341. 

\noindent
{\it Main software:}
{\tt astropy} \cite{2013ascl.soft04002G};
{\tt bilby} \cite{2019ApJS..241...27A};
{\tt cBHBd} \cite{2020MNRAS.492.2936A};
{\tt jax} \cite{jax2018github};
{\tt numpy} \cite{2020Natur.585..357H};
{\tt numpyro} \cite{phan2019composable,bingham2019pyro};
{\tt scipy} \cite{2020SciPy-NMeth}.
\\

\noindent{\it Data and code availability}. The cluster and hierarchical inference codes used to produce the results in this paper and the resulting data products are available under reasonable request.
\end{acknowledgments}

\clearpage

\setcounter{equation}{0}
\setcounter{figure}{0}
\setcounter{table}{0}

\renewcommand{\theequation}{S\arabic{equation}}
\renewcommand{\thefigure}{S\arabic{figure}}
\renewcommand{\thetable}{S\arabic{table}}

\setcounter{page}{1}
\appendix
\onecolumngrid

\begin{table}
\begin{center}
 \renewcommand{\arraystretch}{1.05}
 \begin{tabular}{c l l}
 \hline
 \hline
 Parameter & Prior & Defined in \\
 \hline
  $a_{\chi}$
    & $\mathcal{HN}(3)$ & equation~\eqref{Xeff_GP}  Model 1\\
    $\ln \ell_{\chi}$
    & $\mathcal{N}(-0.5,1)$ & equation~\eqref{Xeff_GP}  Model 1\\
  $a_{\chi}^2$
    & $\mathcal{LU}(0,2)$ & equation~\eqref{Xeff_GP} Model 2\\
    $\ell_{\chi}$
    & $\mathcal{LU}(-1,1)$ & equation~\eqref{Xeff_GP} Model 2\\
 \hline
$a_{m}$ &$\mathcal{HN}(3)$ & equation~\eqref{massGP} Model 1\\
  $\ln \ell_{m}$  & $\mathcal{N}(0,1)$ & equation~\eqref{massGP} Model 1\\
 $a_{m}^2$ & $\mathcal{LU}(0,2)$ & equation~\eqref{massGP} Model 2\\
  $\ell_{m}$  & $\mathcal{LU}(-1,1)$ & equation~\eqref{massGP} Model 2\\
 $a_{\zeta}$, $a_{\mu}$, $a_\sigma$
    & $\mathcal{HN}(4)$ & equations~\eqref{kzeta}, \eqref{mu_sig_m} \\
  $\ln \ell_{\zeta}$, $\ln \ell_{\mu}$, $\ln \ell_\sigma$
    & $\mathcal{N}(-0.5,1)$ & equations~\eqref{kzeta}, \eqref{mu_sig_m} \\
     $a_{\beta_q}$
    & $\mathcal{HN}(5)$ & equation~\eqref{bq_gp}  \\
    $\ln \ell_{\beta_q}$
    & $\mathcal{N}(-0,5,1)$ & equation~\eqref{bq_gp}  \\
  \hline
 $\tilde{m}$
    & $\mathcal{U}(20,100)$ & equation~\eqref{UL_ind}  \\
    $\mu$  
    & $\mathcal{U}(-1,1)$ &  equation~\eqref{eq:pi1}\\
    $\sigma$   
    & $\mathcal{LU}(-1.5,0)$ &  equation~\eqref{eq:pi1}
    \\
      $\chi_{\rm eff,max}$  
    & $\mathcal{U}(0.05,1)$ &  equation~\eqref{UL_ind}
    \\
     $\chi_{\rm eff,min}$  
    & $\mathcal{U}(-1,\chi_{\rm eff,max})$ &  equation~\eqref{UL_ind}
    \\
$\beta_q$ & $\mathcal{N}(0,3)$ & equation~\eqref{eq:pm2} \\
$\kappa$ & $\mathcal{N}(0,6)$ & equation~\eqref{eq:pz} \\
 \hline
 \hline
\end{tabular}
\caption{
Priors adopted for the hyperparameters with which we describe the primary mass, mass ratio, and redshift distributions of the BBH population.
}
\label{tab:priors}
\end{center}
\end{table}

\begin{figure}
\includegraphics[width=.48\columnwidth]{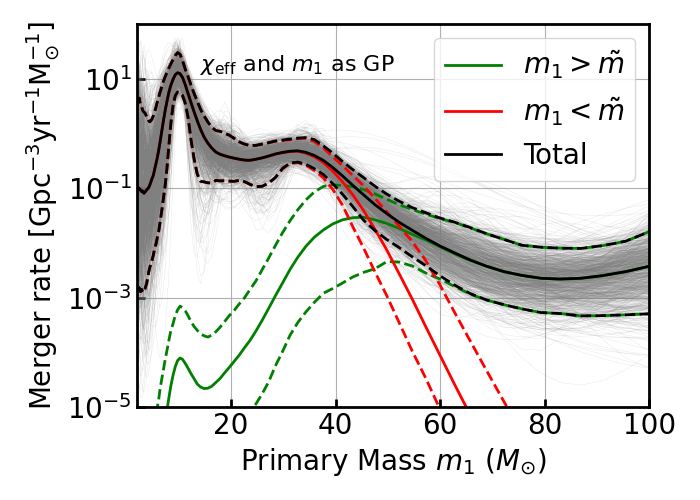} 
\includegraphics[width=.48\columnwidth]{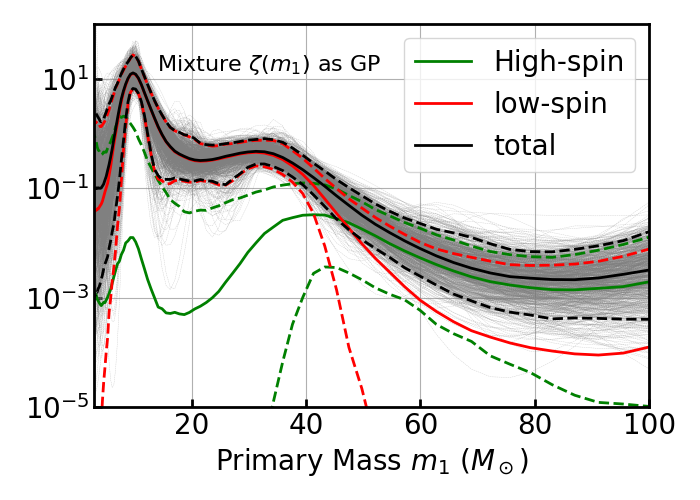} 
    \caption{The differential merger rate as a
function of primary mass. In the left  panel we use the  flexible representation of the $\chi_{\rm eff}$ distribution given in  equation~(\ref{Xeff_GP}).
In the right panel we show the merger rate inferred under the model of equation~\eqref{eq:pi1} where the $\chi_{\rm eff}$ is a mixture between a low-spin and a high spin  populations, with a mass dependent mixture fraction $\zeta(m_1)$. In both cases the data favour a transition to a different population above $\simeq 45M_\odot$, although
the mass value at which this occurs becomes more uncertain under the latter model.
}
\label{fig:massplot}
\end{figure}

\begin{figure} 
\includegraphics[width=0.5\columnwidth]{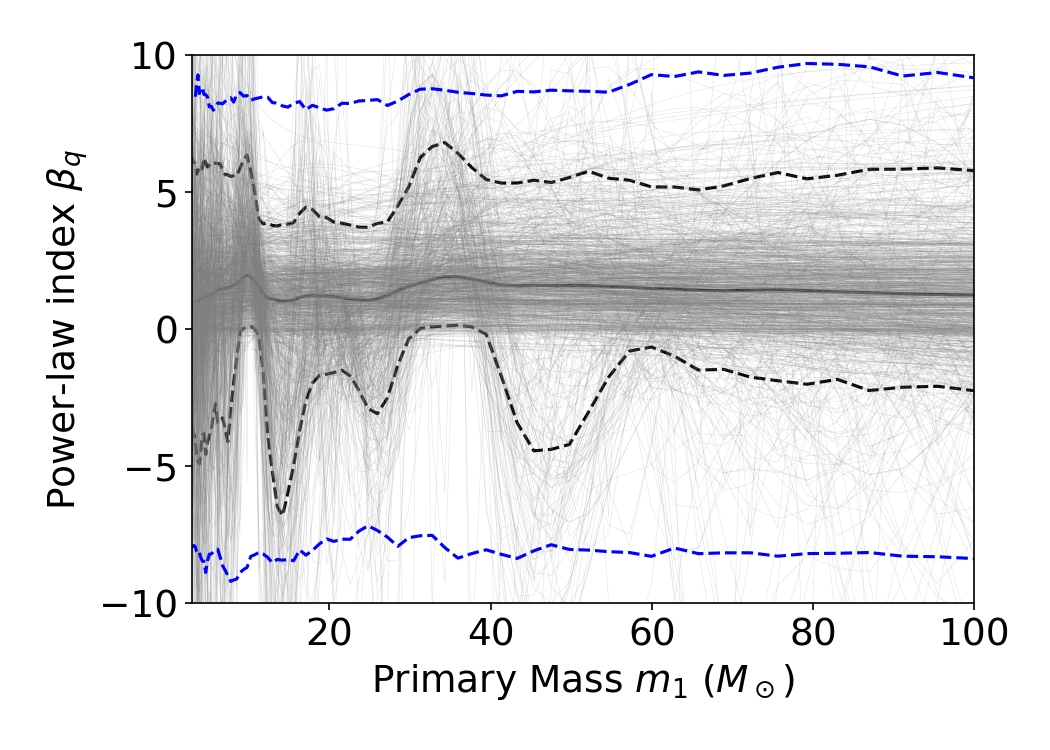} 
    \caption{Power-law index of the mass-ratio distribution
    evaluated using  a GP over a uniform grid in $\log m_1$. For reference, the blue-dashed lines show the $90\%$ confidence intervals of the prior.
    }
    \label{fig:betaq}
\end{figure}

\section{Mass distribution}\label{mass}
In Fig.~\ref{fig:massplot}, we present the primary mass distribution inferred under our suite of models. The recovered features are consistent with previous findings~\cite{LVKCollab2023}, revealing prominent peaks in the differential merger rate around \(10\,M_\odot\) and \(35\,M_\odot\). These structures align with earlier studies that attributed such features to underlying astrophysical imprints in the BH mass function~\cite{2021ApJ...913L..19T,2025A&A...694A.186G}.

For systems with \(m_1 \lesssim 40\,M_\odot\), the population is well described by a single truncated Gaussian effective spin distribution. In this lower-mass regime, the contribution from the isotropic spin component remains minimal.

When the model includes a mass-dependent transition between two spin populations, the inference robustly recovers a sharp change in the \(\chi_{\rm eff}\) distribution at a characteristic mass of \(\tilde{m} \simeq 45\,M_\odot\). This is reflected in the narrow posterior for \(\tilde{m}\), as shown in Fig.~\ref{fig:tmass}. Importantly, this transition mass appears to be largely insensitive to the specific functional form used to model the primary mass distribution.

Furthermore, even when using the  non-parametric model described in Section~\ref{mix}, which allows the mixture fraction \(\zeta(m_1)\) between spin subpopulations to vary with mass, the data continue to favour a distinct transition to a highly spinning subpopulation near the same mass scale. However, in this case, the exact fraction of the isotropic component at high masses is less well constrained. This is likely due to the increased flexibility of the model and the limited number of high-mass mergers in the current dataset. Nonetheless, the posterior distributions still require a non-zero contribution from the isotropic population at \(m_1 \gtrsim 45\,M_\odot\).

Finally, we note that the increased variance in the inferred mass distribution at very low (\(\lesssim 5\,M_\odot\)) and very high (\(\gtrsim 100\,M_\odot\)) masses reflects a reversion to the prior in regions where observational constraints are weak.

\section{Mass ratio $vs$ primary mass}\label{qvsm}

Mergers involving a BH that is itself the product of a previous merger are expected to form a distinct population compared to mergers involving first generation BHs only.
Not only should this population exhibit larger primary masses and spins, it should additionally be characterized by a distinct mass-ratio distribution preferentially favouring unequal-mass binaries.

If the distinct BH spin distributions described above are indeed indicative of a high-mass population of second-generation BHs, then we additionally expect to see signs of a mass ratio distribution that similarly varies with primary mass.
To search for evidence of such behaviour, we extend the power-law model [Eq.~\eqref{eq:pm2}] of secondary masses, promoting the power-law index $\beta_q$ from a constant to itself be a function of primary mass: $\beta_q(m_1)$.
We model \(\beta_q(m_1)\) as
\begin{equation}
\label{bq_gp}
\beta_q(m_1) = \Xi[\ln(m_1)],
\end{equation}
where, as before, \(\Xi\) is drawn from a GP prior evaluated on a regular grid in $\log m_1$
and we use an exponential kernel with variance and smoothing length $a_{\beta_q}$ and $\ell_{\beta_q}$, respectively.
This flexible approach allows us to explore whether the distribution of  mass ratio has a detectable shift above a characteristic mass scale.

We present the results of this analysis in Figure~\ref{fig:betaq}.
There are indications of possible mass-dependence of $\beta_q(m_1)$.
In particular, we see that data constrain $\beta_q\geq 0$ at both $10\,M_\odot$ and $35\,M_\odot$ at $95\%$ credibility, whereas $\beta_q$ is permitted to be smaller or negative between approximately $15$--$30\,M_\odot$ and above $40\,M_\odot$.
However, the data are not sufficiently informative to determine robustly whether the mass-ratio distribution exhibits systematic trends or transitions;
the results above may be due simply to the elevated uncertainties in the $15$--$30\,M_\odot$ and $40$--$100\,M_\odot$ ranges due to the small number of events observed with these masses.
\section{Hierarchical Bayesian Inference}
\label{app:inference-methods}
We carry out hierarchical inference on the BBH population using Hamiltonian Monte Carlo (HMC), as implemented in \texttt{numpyro}, a probabilistic programming framework built on \texttt{jax}. HMC requires the likelihood to be a differentiable function of the population hyperparameters. However, the piecewise-defined spin models we use (e.g., equation~\eqref{Xeff_GP}) introduce discontinuities at $m_1 = \tilde{m}$, which result in non-differentiable behavior in the likelihood with respect to $\tilde{m}$. To address this issue, we replace these discontinuous transitions with smooth but sharp interpolations between the functional forms of the effective spin distribution above and below $\tilde{m}$, ensuring differentiability while preserving the essential structure of the model. We redirect the reader to \cite{2025PhRvL.134a1401A} for the details of how this is implemented.

We conduct our analysis within the framework of standard hierarchical Bayesian inference. For each gravitational-wave event, let \( p(\theta_i | d_i) \) denote the posterior distribution over the source parameters \( \theta_i \) (such as component masses, redshift, and spins), conditioned on the observed data \( d_i \). The posterior on the population hyperparameters \( \Lambda \), given data from all detected events, is then given by~\cite[e.g.,][]{
Fishbach_2018,2019MNRAS.486.1086M,2022ApJ...937L..13C}
\begin{equation}
    p(\Lambda \,|\, \{d_i\})
        \propto p(\Lambda)\, \xi^{-N_{\rm obs}}(\Lambda)
        \prod_{i=1}^{N_{\rm obs}}
        \int d\theta_i\, p(\theta_i|d_i)
        \frac{p(\theta_i|\Lambda)}{p_\mathrm{pe}(\theta_i)}\ ,
    \label{eq:likelihood-integral}
\end{equation}
where \( p_\mathrm{pe}(\theta_i) \) is the prior used in the original parameter estimation analysis, and \( p(\Lambda) \) is our prior on the hyperparameters governing the population distribution. 
Rather than performing the full integral over each posterior, we approximate the expectation using discrete posterior samples for each event. This yields:
\begin{equation}
    p(\Lambda \,|\, \{d_i\})
        \propto p(\Lambda)\,\xi^{-N_{\rm obs}}(\Lambda)
        \prod_{i=1}^{N_{\rm obs}}
        \bigg\langle
        \frac{p(\theta_i|\Lambda)}{p_\mathrm{pe}(\theta_i)}
        \bigg\rangle,
    \label{eq:likelihood-sum}
\end{equation}
where the angle brackets denote an average over the posterior samples of each event.

The detection efficiency \( \xi(\Lambda) \) accounts for the fraction of astrophysical signals that we expect to detect under the population model \( \Lambda \). It is estimated using injection campaigns consisting of \( N_{\rm inj} \) simulated signals drawn from a reference distribution \( p_\mathrm{inj}(\theta_i) \). Considering only the \( N_\mathrm{found} \) injections that are successfully recovered by at least one search pipeline with a false alarm rate below 1~yr\(^{-1} \), we compute:
\begin{equation}
    \xi(\Lambda) = \frac{1}{N_{\rm inj}} \sum_{i=1}^{N_{\rm found}} \frac{p(\theta_i | \Lambda)}{p_\mathrm{inj}(\theta_i)}\,,
    \label{csi}
\end{equation}
where the ratio reweights the found injections from the injection prior to the proposed population model. This approach follows the methodology described in~\cite{2022ApJ...937L..13C}, and uses the injection campaign presented in~\cite{2024PhRvD.109b2001A,injections}.

Finally, to guard against potential biases from finite sampling effects in the evaluation of Eq.~\ref{eq:likelihood-sum}, we compute the number of effective samples \( N_{\mathrm{eff}} \) that contribute to the Monte Carlo estimate of the likelihood for each gravitational-wave event. Given a set of \( N_i \) posterior samples \( \{\theta_{i,j}\}_{j=1}^{N_i} \) for event \( i \), the effective sample count under a proposed population model \( \Lambda \) is defined as
\begin{equation}
    N_{\mathrm{eff},i}(\Lambda) \equiv \frac{\left[ \sum_{j=1}^{N_i} w_{i,j}(\Lambda)\right]^2}{\sum_{j=1}^{N_i} \left[ w_{i,j}(\Lambda)\right]^2}\,,
    \label{eq:Neff}
\end{equation}
where the weights \( w_{i,j}(\Lambda) = p(\theta_{i,j}|\Lambda)/p_\mathrm{pe}(\theta_{i,j}) \) reweight each sample according to the proposed population model.
A small value \( N_{\mathrm{eff},i}(\Lambda) \lesssim 10 \) indicates that only a handful of posterior samples are informative under the model \( \Lambda \), which may lead to increased sensitivity to Monte Carlo noise \cite{2022arXiv220400461E}.

Similarly, the effective number of injections contributing to the calculation of
Eq.~\ref{csi} is defined as
\begin{equation}
N_\mathrm{eff}^\mathrm{inj}(\Lambda) = \frac{\left(\sum_i w_i(\Lambda)\right)^2}{\sum_j w_j^2(\Lambda)}.
\end{equation}
To ensure that the systematic uncertainty in \( N_\mathrm{exp}(\Lambda) \) remains a subdominant effect in our hierarchical analysis, it is necessary that
$N_\mathrm{eff}^\mathrm{inj}(\Lambda) \gtrsim 4 N_\mathrm{obs}$ \cite{2022arXiv220400461E}.

We track the statistic \( \mathcal{N} \equiv \min \log \left[ N_{\mathrm{eff},i}(\Lambda) \right] \) 
and compute $N_\mathrm{eff}^\mathrm{inj}$
for each population model.
 We then safeguard the inference by strongly penalizing models for which \( N_\mathrm{eff}^\mathrm{inj} < 4N_\mathrm{obs} \) and/or \( \min \log N_\mathrm{eff}^\mathrm{} < 0.6 \).
Specifically, we define the function
\begin{equation}
S(x) = \frac{1}{1 + x^{-30}},
\end{equation}
which approaches unity for large \( x \) and falls rapidly to zero as \( x \) approaches zero. We then include the terms
\begin{equation}
\ln S\left(\frac{N_\mathrm{eff}^\mathrm{inj}}{4N_\mathrm{obs}}\right) + \ln S\left(\frac{ \mathcal{N}}{0.6}\right)
\end{equation}
in the log-likelihood calculation implemented in NumPyro. This ensures the log-likelihood diverges to \( -\infty \) if either condition is violated, thereby excluding models with pathologically low effective sample sizes.

\section{Priors and hyperparameter posteriors}\label{priors}
\begin{figure*}
    \centering  
    \includegraphics[width=1.\textwidth]{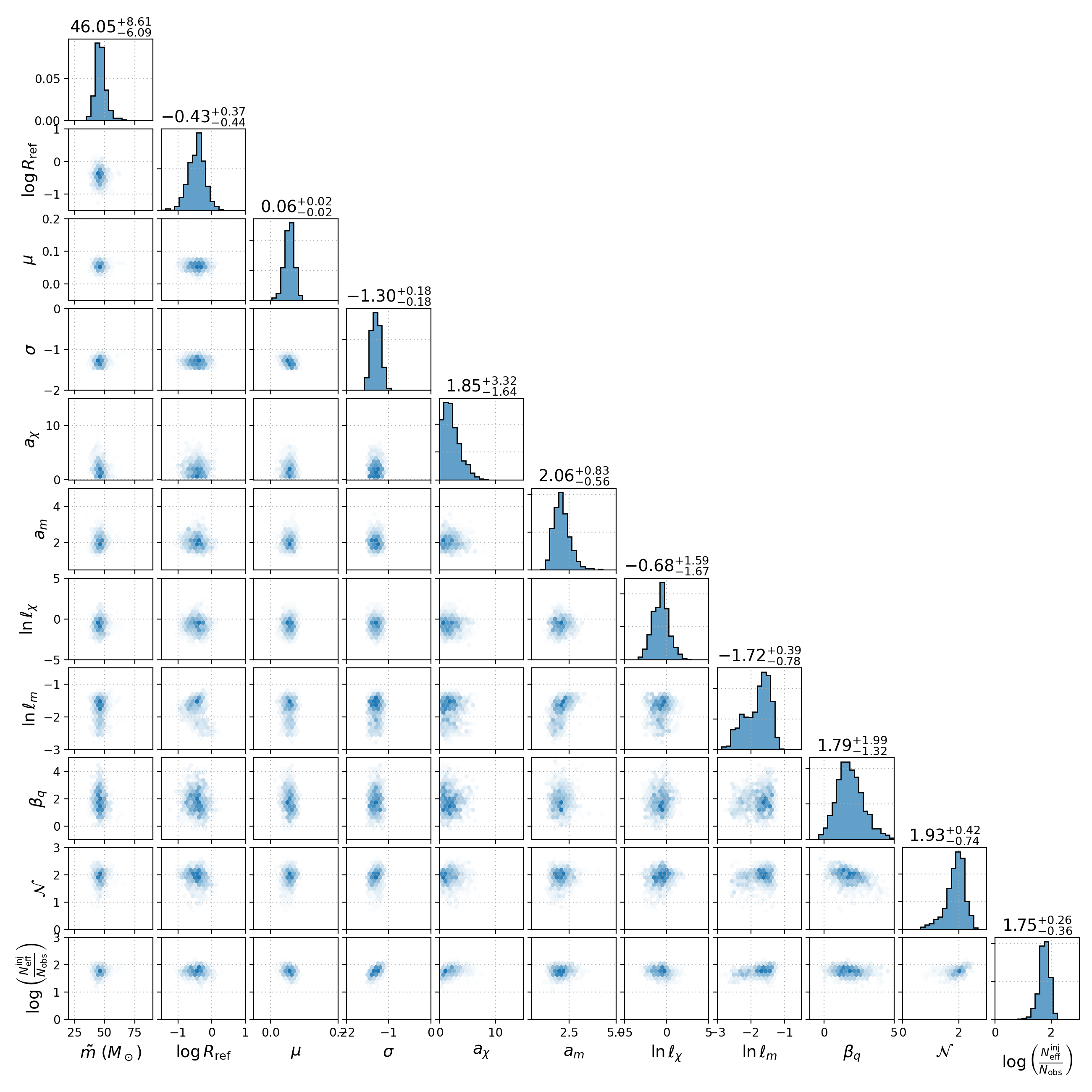}
    \caption{Posteriors on the parameters that govern the hierarchical model where
    the $\chi_{\rm eff}$ distribution is
    represented by 
    a fixed Gaussian 
    below $\tilde{m}$
    and a non-parametric distribution above $\tilde{m}$. 
    In this case the priors on the kernel parameters
    of the GP prior corresponding to Model 1 in Fig.\ \ref{fig:XeffD} and Table\ \ref{tab:priors}.
    Here $N_{\rm eff}^{\rm inj}/N_{\rm obs}$ gives the total number of injection divided by the number of  detections. $R_{\rm ref}$ represents the differential merger rate
    value in units of $\rm Gpc^{-3}yr^{-1}M_{\odot}^{-1}$ evaluated at $20M_\odot$ and at redshift $z=0.2$.
    }
    \label{fig:post2}
\end{figure*}

\begin{figure*}
    \centering  
    \includegraphics[width=1.\textwidth]{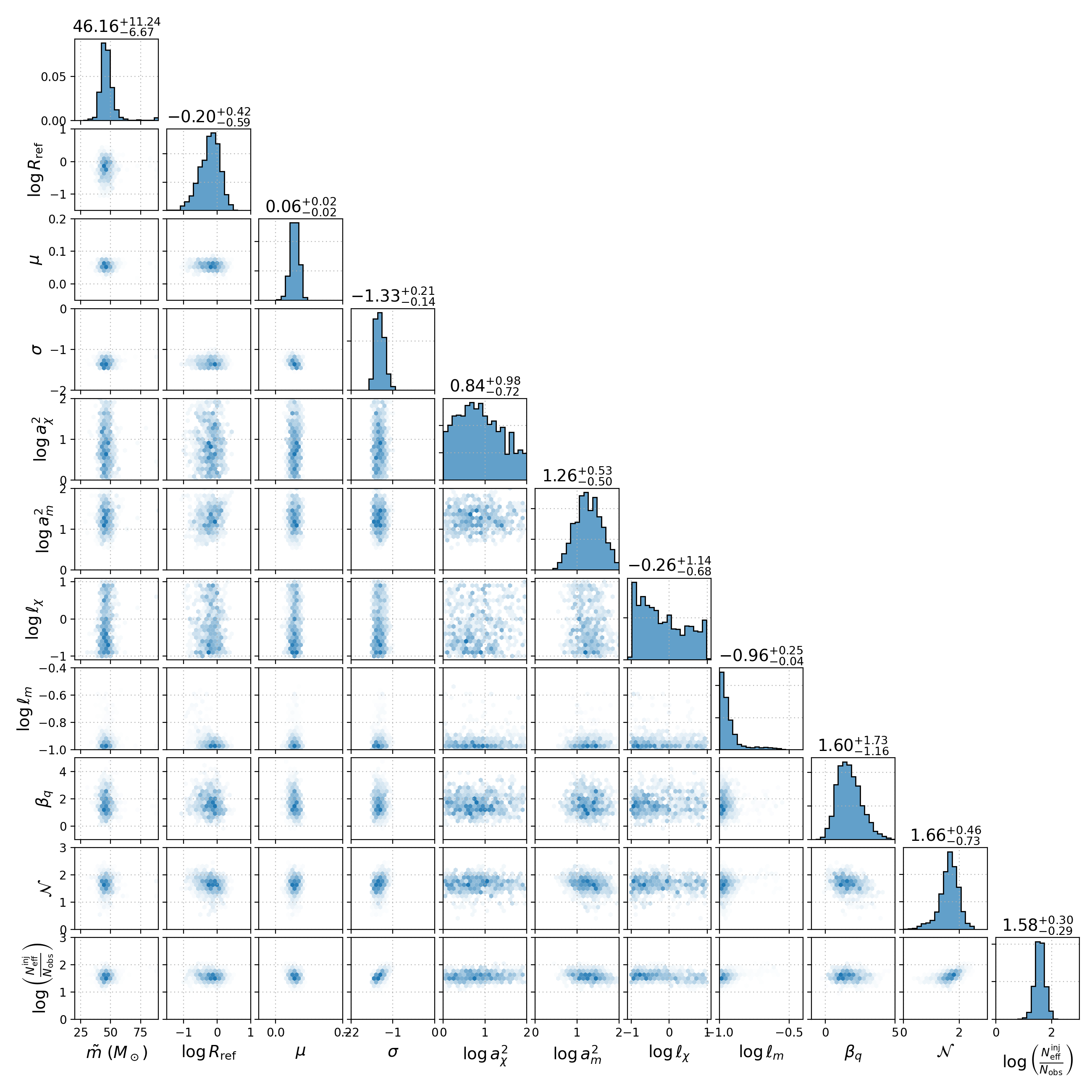}
    \caption{Posteriors on the parameters that govern the hierarchical model where
    the $\chi_{\rm eff}$ distribution is
    represented by 
    a fixed Gaussian 
    below $\tilde{m}$
    and a non-parametric distribution above $\tilde{m}$. 
    In this case the priors on the kernel parameters
    of the GP prior corresponding to Model 2 in Fig.\ \ref{fig:XeffD} and Table\ \ref{tab:priors}.
    Here $N_{\rm eff}^{\rm inj}/N_{\rm obs}$ gives the total number of injection divided by the number of  detections. $R_{\rm ref}$ represents the differential merger rate
    value in units of $\rm Gpc^{-3}yr^{-1}M_{\odot}^{-1}$ evaluated at $20M_\odot$ and at redshift $z=0.2$.
    }
    \label{fig:post1}
\end{figure*}

\begin{figure*}
    \centering  
    \includegraphics[width=1.\textwidth]{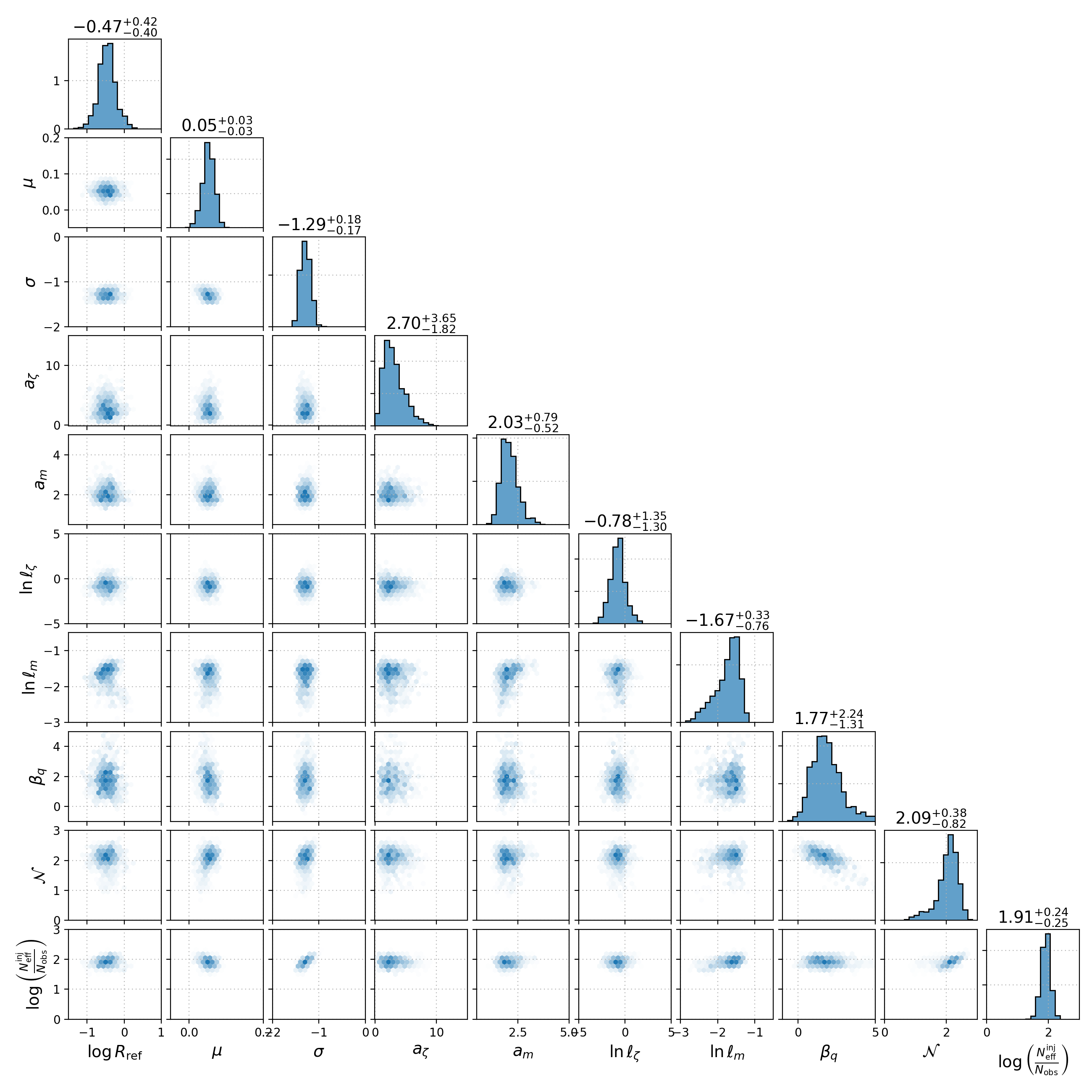}
    \caption{Posteriors on the key parameters that govern the hierarchical model of Eq.~\eqref{eq:pi1} where the $\chi_{\rm eff}$ distribution is represented as 
     mixture  between a Gaussian and a uniform distribution with the mixing fraction modeled as a GP over $\log m_1$. 
    Here $N_{\rm eff}^{\rm inj}/N_{\rm obs}$ gives the total number of injection divided by the number of  detections. $R_{\rm ref}$ represents the differential merger rate
    value in units of $\rm Gpc^{-3}yr^{-1}M_{\odot}^{-1}$ evaluated at $20M_\odot$ and at redshift $z=0.2$.
    }
    \label{fig:post3}
\end{figure*}

\begin{figure*}
    \centering  
    \includegraphics[width=1.\textwidth]{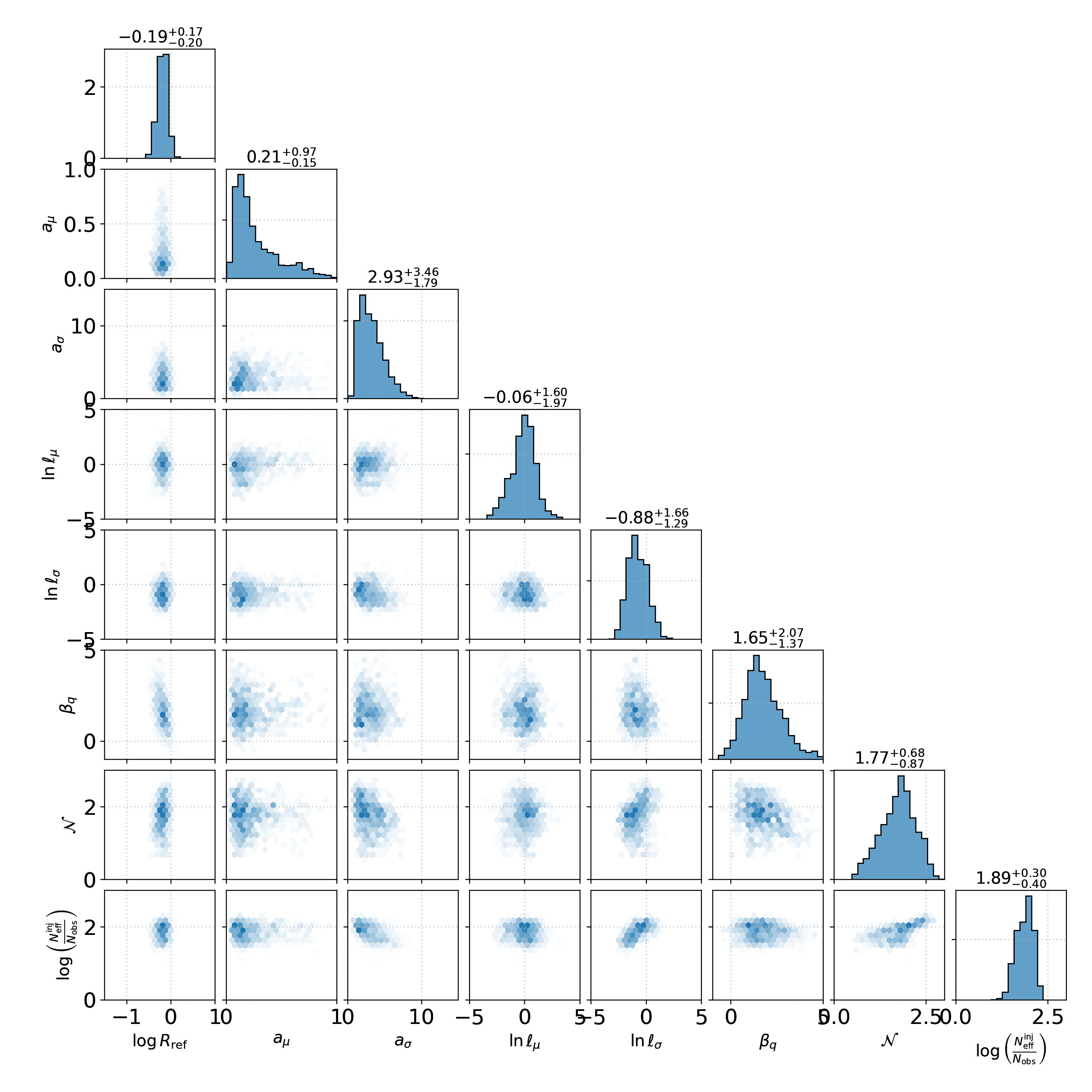}
    \caption{Posteriors on the key  parameters that govern the hierarchical model of Section~\ref{musigma} 
 where the $\chi_{\rm eff}$ distribution is represented as 
     truncated Gaussian with mass dependent mean and variance, both modeled non-parametrically as a GP. 
    Here $N_{\rm eff}^{\rm inj}/N_{\rm obs}$ gives the total number of injection divided by the number of  detections. $R_{\rm ref}$ represents the differential merger rate
    value in units of $\rm Gpc^{-3}yr^{-1}M_{\odot}^{-1}$ evaluated at $20M_\odot$ and at redshift $z=0.2$.
    }
    \label{fig:post4}
\end{figure*}

\begin{figure*}
    \centering   \includegraphics[width=1.\textwidth]{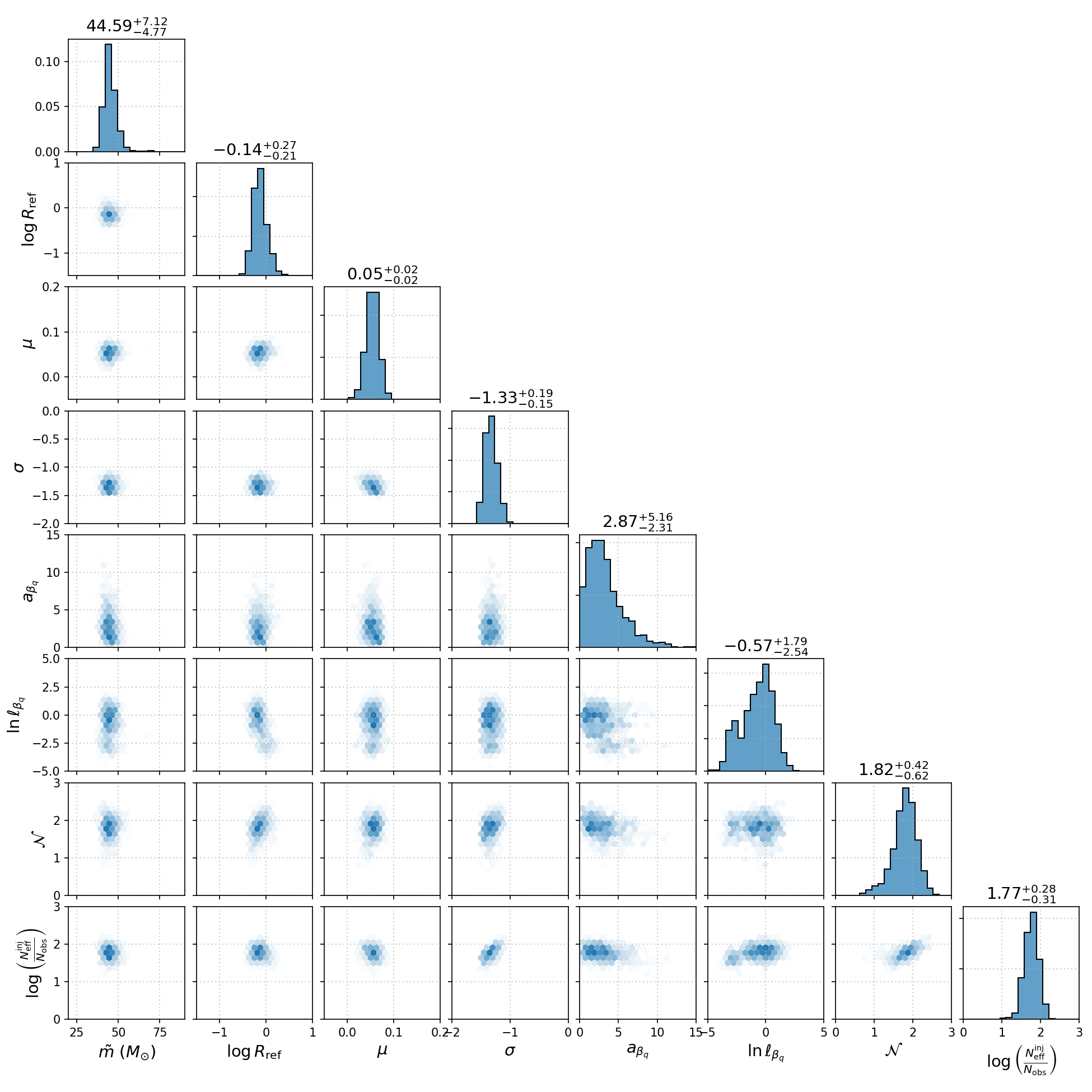}
    \caption{Posteriors on the key  parameters that govern the hierarchical model of Appendix~\ref{qvsm} with a mass dependent
 power-law index of the mass ratio distribution, which is represented non-parametrically using 
     a GP over a mass grid. 
    Here $N_{\rm eff}^{\rm inj}/N_{\rm obs}$ gives the total number of injection divided by the number of  detections. $R_{\rm ref}$ represents the differential merger rate
    value in units of $\rm Gpc^{-3}yr^{-1}M_{\odot}^{-1}$ evaluated at $20M_\odot$ and at redshift $z=0.2$.
    }
    \label{fig:post5}
\end{figure*}

The specific priors adopted for our analysis are summarized in Table~\ref{tab:priors}. Figg.~\ref{fig:post2}, ~\ref{fig:post1},~\ref{fig:post3},~\ref{fig:post4}, and~\ref{fig:post5} show the posteriors for the key parameters of our models.

As shown in Section~\ref{npm}, certain features of the inferred posteriors are sensitive to the choice of priors for the kernel parameters, which govern the behaviour of the latent function in the Gaussian Process. This kind of sensitivity is typical in non-parametric models, where one must decide how smooth the function should be across the grid, and where the posterior can revert to the prior in regions of the GP grid where the data are not informative.

To address this, we ran our models multiple times with different prior choices and selected those that allowed for  flexibility and broad variation in the latent function, while still yielding accurate and stable likelihood estimates. 
In Model 1, we adopt normal distributions for the priors on the kernel amplitude $a_{\chi}$ and the logarithm of the length scale $\ln \ell_{\chi}$, following~\cite{2024PhRvX..14b1005C}. Specifically, we take
$
a_{\chi} \sim \mathcal{N}(0,\,3)$ and $\ln \ell_{\chi} \sim \mathcal{N}(-1,\,0.5)$,
applied over the $\chi_{\mathrm{eff}}$ grid.

These prior choices are quantitatively motivated as follows:
the maximum fractional variation we physically expect in the merger rate across the grid is of order $\sim 10^3$. The grid size  is $\Delta x = 2$.
For a stationary GP with a squared exponential kernel and amplitude $a_{\chi}$, the prior typical fluctuation (in $\log$ rate) between two points separated by a distance $\Delta x$ is
$
\mathrm{Std}[\Delta f] = \sqrt{2} a_{\chi} \left( 1 - \exp\left[ -\frac{(\Delta x)^2}{2\ell_{\chi}^2} \right] \right)^{1/2}.
$
 For large $\Delta x/\ell_{\chi}$, this saturates to $\sqrt{2} a_{\chi}$.
Setting $a_{\chi} = 3$, we expect a prior standard deviation in $\log$ rate of about $4.2$ (since $\sqrt{2} \times 3 \approx 4.2$), comfortably covering the expected physical variation of up to three orders of magnitude, but without being overwhelmingly broad.
The prior on $\ell_{\chi}$ centers the length scale at $\exp(-1) \approx 0.37$ (with most prior weight between about $0.2$ and $0.6$), so that the GP can model features ranging from fairly localized to nearly flat across the grid, but discourages extremely rapid oscillations or excessively global trends unless strongly favored by the data.
We find that  these choices provide a balance between flexibility and regularization, and ensure that the GP  adapts to physically plausible variations while penalizing models that are too rigid. 

Similar prior choices were adopted for the remaining models, and can be similarly justified. In Model 2, we adopted log-uniform priors on the kernel parameters \(a^2\)  and \(\ell\) to model both the $\chi_{\rm eff}$ and $m_1$ distributions.
We set a lower bound on 
$ \ell$ to avoid short-scale variations that would result in small, unphysical features and poor sampling efficiency. The upper bound is fixed at \(\log \ell = 2\), which is significantly larger than the grid full width \(\Delta x = 2\) used for both \(\log m_1\) and \(\chi_{\rm eff}\).
To allow for large but physically plausible variations in the log rate, we impose an upper bound which is again derived from the maximum expected variation across the parameter space, \(\delta_{\rm max} R \sim 1000\). This requires \(a^2 \gtrsim \ln(1000)^2 \sim 50\) to ensure that the GP prior is flexible enough to capture significant features in the data, while maintaining numerical stability and well-behaved posteriors.


\newpage

\end{document}